\documentclass[11pt,a4paper]{article}

\usepackage{jheppub}

\usepackage{natbib}
\usepackage{amsfonts,amsmath,amssymb}
\usepackage{hyperref}
\usepackage[toc,page]{appendix}
\usepackage{graphicx}

\def\bra#1{\mathinner{\langle{#1}|}}
\def\ket#1{\mathinner{|{#1}\rangle}}
\def\expect#1{\langle#1\rangle}

\def\ol#1{\bar{#1}}

\def\ua{{\uparrow}}
\def\da{{\downarrow}}

\def\lra{{\longrightarrow}}

\def\pd{\partial}

\newcommand{\ii}{{\rm i}}
\newcommand{\dd}{{\rm d}}

\newcommand{\ZZp}{\mathbb{Z}_{\geq 0}}
\newcommand{\RaR}{\mathbb{R}}
\newcommand{\CC}{\mathbb{C}}

\def\Re{{\,{\rm Re}\,}}

\def\squote#1{\lq{#1}\rq}

\def\rhoh{\ol{\rho}}

\def\frakn{{\mathfrak{n}}}
\def\frakh{{\ol{\mathfrak{n}}}}
\def\frakI{{\mathfrak{I}}}
\def\fraka{{\mathfrak{a}}}

\newcommand{\T}{{T}}
\newcommand{\Q}{{Q}}
\newcommand{\XX}{X}
\newcommand{\D}{G}
\newcommand{\len}{N}
\newcommand{\kd}{\boldsymbol{\delta}}

\usepackage[parfill]{parskip}
\begin{document}

\title{String-charge duality in integrable lattice models}
\author[1]{Enej Ilievski, }
\author[1]{Eoin Quinn, }
\author[2]{Jacopo De Nardis, }
\author[3]{and Michael Brockmann}

\affiliation[1]{Institute for Theoretical Physics,
University of Amsterdam, Science Park 904, Postbus 94485, 1090 GL Amsterdam, The Netherlands}
\affiliation[2]{Laboratoire de Physique Th\'eorique, Ecole Normale Sup\'erieure, Paris, France}
\affiliation[3]{Max Planck Institute for the Physics of Complex Systems, N\"othnitzer Str.~38, 01187 Dresden, Germany}
\emailAdd{e.ilievski@uva.nl}

\arxivnumber{1512.04454}

\date{\today}

\abstract{
We present an identification of the spectra of local conserved operators of integrable quantum lattice models and the density 
distributions of their thermodynamic particle content. This is derived explicitly for the Heisenberg XXZ spin chain.
As an application we discuss a quantum quench scenario, in both the gapped and critical regimes.
We outline an exact technique which allows for an efficient implementation on periodic matrix product states.
In addition, for certain simple product states we obtain closed-form expressions for the density distributions
in terms of solutions to Hirota difference equations. Remarkably, no reference to a maximal entropy principle is invoked.
}

\maketitle
\flushbottom

\clearpage
\section{Introduction}

Over recent years, one of the main challenges in the theory of low-dimensional quantum many-body systems has been 
to understand the relaxation and equilibration of local observables.
In particular, a central question has been to explain the emergence of statistical ensembles
in isolated systems. For systems with generic interactions, one can resort to the argument 
of typicality, which is customarily formulated in the view of the Eigenstate Thermalization Hypothesis~\cite{Deutsch91,Srednicki94},
stating that eigenfunctions with similar energy densities cannot be distinguished on the level of local correlators,
and that the behaviour in the long-time limit should agree with the predictions of the canonical Gibbs ensemble,
with temperature fixed by the initial energy density. Exceptions to this paradigm are non-interacting models and integrable 
interacting quantum many-body problems, for which this picture breaks down due to the existence of a macroscopic number of charges, 
i.e.~conserved local operators.

A recent study on the anisotropic Heisenberg spin chain has shown that there exist sufficient local conservation laws to 
completely determine the equilibrium steady state after a quantum quench~\cite{Ilievski_GGE}.
This followed an observation~\cite{IMP15} that although these conservation laws are comprised of a tower of charges which
are functionally related, they are nevertheless linearly independent, and represent an adequate vector space for description of 
equilibrium ensembles.

In this paper we clarify this relationship and present a direct identification between the steady state of an integrable system and 
its charges, see Eq.~\eqref{eqn:Laplace_rho_X_gapped}, expressed as a discrete wave equation.
The conceptual consequence of this result is that the conventional
perspective~\cite{RigolNature,Rigol11,Polkovnikov_colloquium,Gogolin_review,Langen15,Ilievski_GGE}, based on
generalized free energy functionals and a maximal entropy principle, is not a necessary framework for the characterization
of equilibrium states in quantum integrable lattice models.

More specifically, we examine quantum quenches in the full range of the anisotropic Heisenberg spin chain, restricting to the 
properties of the steady state. We cast our discussion in the language of fusion hierarchies, which is the universal language of 
integrability, and renders manifest the locality of the charges. Special emphasis is devoted to the case of the quantum critical 
(gapless) regime where several interesting exceptional features arise.
Moreover, we re-derive exact results for the steady state for quenches for both the N\'eel and dimer 
states~\cite{Brockmann14,Mestyan15}, as closed-form solutions to the quantum Hirota difference equation~\cite{KLWZ97,Zabrodin97}.

The paper is organised as follows. In Sec.~\ref{sec:generalities} we introduce the technical tools at our disposal: Lax and transfer 
operators, Hirota equation and Y-system, Baxter's Q-operator, and  string densities. The charges are defined
in Sec.~\ref{sec:charge-string}, and their relationship to the string densities is derived. In Sec.~\ref{sec:charge_densities} we 
discuss the quantum quench scenario, describe the evaluation of the charges on initial states, and present exact results for quenches 
from both the dimer and N\'eel states. We summarize our presentation is Sec.~\ref{sec:conclusions}, and discuss open ends. We focus 
our discussion on the gapped regime, and present the modifications necessary for the critical regime in subsections and appendices.

\section{Anisotropic Heisenberg spin-1/2 chain}
\label{sec:generalities}

In this paper we focus on the Heisenberg XXZ Hamiltonian, which on a periodic chain of $\len$ sites reads
\begin{equation}
H=\sum_{n=1}^{\len}\sigma^{x}_{n}\sigma^{x}_{n+1}+\sigma^{y}_{n}\sigma^{y}_{n+1}
+\Delta (\sigma^{z}_{n}\sigma^{z}_{n+1}-1),
\label{eqn:Hamiltonian}
\end{equation}
written in the basis of Pauli matrices $\sigma^{\alpha}$ for $\alpha\in\{x,y,z\}$.
The Hamiltonian density has a $U(1)$ symmetry associated with a local conserved operator $S^{z}=\sum_{n}s^{z}_{n}$.
Anisotropy of the interaction is controlled by parameter $\Delta \in \RaR$ conventionally parametrized
by $q$-deformation parameter, $\Delta=\tfrac{1}{2}(q+q^{-1})$.
There are two regimes which are to be distinguished, meeting at the isotropic point $\Delta=1$: (i)
\emph{gapped regime} with $q=e^{\eta}$, $\eta >0$, and (ii) \emph{gapless regime} corresponding to the interval
$|\Delta|\leq 1$, where $q$-parameter takes values on the unit circle, $q=\exp{(\ii \gamma)}$ for $\gamma \in [0,\pi)$.
In the following we focus our discussion on the gapped regime. Necessary modifications
required to treat the gapless regime are discussed in Sec.~\ref{sec:gapless}.

\subsection{Lax and transfer operators}
\label{sec:Lax}

The Heisenberg XXZ Hamiltonian belongs to an infinite family of commuting local operators.
The underlying integrable structure is encoded in the $\mathcal{U}_{q}(\mathfrak{su}(2))$-invariant Lax matrix.
We employ a general unitary spin-$\tfrac{j}{2}$ representation over auxiliary spaces $\mathcal{V}_{j}$ of dimension $j+1$, and
define a family of Lax operators on $\mathcal{V}_{1}\otimes \mathcal{V}_{j}$
\begin{eqnarray}
L_{j}(\mu) = \frac{1}{\sinh{(\eta)}}
\begin{pmatrix}
\sin{(\mu+\ii\eta s^{z})} & \ii \sinh{(\eta)}s^{-} \cr
\ii \sinh{(\eta)}s^{+} & \sin{(\mu-\ii\eta s^{z})}
\end{pmatrix},
\label{eqn:Lax_matrix}
\end{eqnarray}
which act as $2\times 2$ matrices over the local spin space $\mathcal{V}_{1}$ with $\mathcal{V}_{j}$-valued components,
and $\mu$ is a complex-valued parameter called the spectral parameter. Spaces $\mathcal{V}_{j}$ are
spanned by a basis $\ket{n}$, with indices $n=0,1,\ldots j$, and where $\ket{0}$ is the highest-weight vector.
Spin operators $s^{\alpha}$ in Eq.~\eqref{eqn:Lax_matrix}, 
with $\alpha = +,-,z$, fulfil the $q$-deformed\footnote{We use the following notation for $q$-numbers,
$[x]_{q}\equiv \sinh{(\eta x)}/\sinh{(\eta)}$.} $\mathfrak{su}(2)$ commutation relations $[s^{+},s^{-}]=[2s^{z}]_{q}$,
$q^{\pm s^{z}}s^{\pm}=q^{\pm 1}s^{\pm}q^{\pm s^{z}}$, and are prescribed as
\begin{eqnarray}
s^{z}\ket{n} &=& (\tfrac{j}{2}-n) \ket{n},\\
s^{-}\ket{n} &=& \sqrt{[j-n]_{q}[n+1]_{q}}\ket{n+1},\\
s^{+}\ket{n+1} &=& \sqrt{[j-n]_{q}[n+1]_{q}}\ket{n}.
\end{eqnarray}

Lax operators serve as local building units for construction of a commuting set of \emph{higher-spin} quantum transfer
operators $T_{j}(\mu)$, for $j\in \ZZp$, using the \squote{traces over monodromies} construction,
\begin{equation}
T_{j}(\mu)={\rm Tr}_{\mathcal{V}_{j}}L^{(1)}_{j}(\mu)L^{(2)}_{j}(\mu)\cdots L^{(\len)}_{j}(\mu),
\label{eqn:T-operators_definition}
\end{equation}
where the superscript indices pertain to embeddings of operators Eq.~\eqref{eqn:Lax_matrix} into the spin-chain Hilbert space
$\mathcal{H}\cong (\CC^{2})^{\otimes \len}$.
Here the trivial (i.e.~$j=0$) representation is the scalar $T_{0}(\mu)=(\sin{(\mu)}/\sinh{(\eta)})^{\len}$.
Commutation $[T_{j}(\mu),T_{j^{\prime}}(\mu^{\prime})]=0$ holds for all $j,j^{\prime}\in \ZZp$ and
$\mu,\mu^{\prime}\in \CC$ by virtue of Yang--Baxter relation~\cite{KorepinBook,Faddeev_arxiv}.

\subsection{Hirota equation and Y-system}
\label{sec:Hirota}

The Hirota difference equation, also known as the $T$-system~\cite{KP92,KNS94,BLZII,KLWZ97}, is a discrete system of
bilinear relations of the form\footnote{We note that the form provided here is not the general one.
For quantum symmetries which belong to higher-rank algebras an additional discrete representation label is required, see 
e.g.~\cite{KNS94}.}
\begin{equation}
T^{+}_{j}T^{-}_{j}=\phi^{[j]}\,\bar{\phi}^{[-j]}+T_{j-1}T_{j+1},\quad j\geq 0,
\label{eqn:Hirota}
\end{equation}
with the boundary condition $T_{-1}\equiv 0$ and the bar denoting complex conjugation. Here and subsequently we
employ a compact notation for representing $k$-unit imaginary shifts,
\begin{equation}
f^{\pm} = f(\mu\pm \tfrac{\ii \eta}{2}),\quad f^{[\pm k]} = f(\mu \pm k\tfrac{\ii \eta}{2}).
\end{equation}
Physically, Eq.~\eqref{eqn:Hirota} encodes the fusion rules of the symmetry algebra of a quantum integrable lattice model.
The sequence of higher-spin transfer operators $T_{j}(\mu)$ defined in Eq.~\eqref{eqn:T-operators_definition} represents
a solution to the Hirota equation with potentials $\phi=T^{+}_{0}$ and $\ol{\phi}= T^{-}_{0}$. We shall refer to it as the 
\emph{canonical} solution~\cite{Gromov09}.

The $T$-system functional hierarchy governing the $T$-operators exhibits a gauge symmetry
\begin{equation}
T_{j}\mapsto g^{[j]}\ol{g}^{[-j]}T_{j},\quad \phi \mapsto g^{+}g^{-}\phi.
\label{eqn:Hirota_gauge}
\end{equation}
The $T$-operators can be combined in a gauge-invariant way as $Y$-operators, related through a non-linear transformation
\begin{equation}
Y_{j}=\frac{T_{j-1}T_{j+1}}{\phi^{[j]}\ol{\phi}^{[-j]}}=\frac{T^{+}_{j}T^{-}_{j}}{\phi^{[j]}\ol{\phi}^{[-j]}}-1.
\label{eqn:Y_to_T_relation}
\end{equation}
$Y$-operators obey the $Y$-system functional hierarchy
\begin{equation}
Y^{+}_{j}Y^{-}_{j}=(1+Y_{j-1})(1+Y_{j+1}),
\label{eqn:Y-system}
\end{equation}
with the boundary condition $Y_{0}\equiv 0$. A convenient way of expressing Eq.~\eqref{eqn:Y_to_T_relation} is
\begin{equation}
\square \log T_{j}=\log\left(1+\frac{1}{Y_{j}}\right),
\end{equation}
where the operator $\square$ is a \emph{discrete d'Alembertian}~\cite{Zabrodin97,KSZ08,KL10} which acts on a discrete family of 
functions $f_{j}$ as
\begin{equation}
\square f_{j}\equiv f^{+}_{j}+f^{-}_{j}-f_{j-1}-f_{j+1}.
\label{eqn:Laplace_gapped}
\end{equation}
In what follows we will consider such $f_{j}$ that are analytic inside the \emph{physical region}
\begin{equation}
\mathcal{P}_{\eta}=\{x\in \CC|\Re(x)\in [-\tfrac{\pi}{2},\tfrac{\pi}{2}],{\rm Im}(x)<\tfrac{\eta}{2}\}.
\label{eqn:physical_region_gapped}
\end{equation}

\subsection{Baxter's $Q$-operator and Bethe equations}
The fundamental transfer matrix $T_{1}$ admits a useful decomposition in terms of the so-called
Baxter's $Q$-operator~\cite{Baxter73,BLZII,shortcut}
\begin{equation}
T_{1}Q=T^{+}_{0}\ol{Q}^{[-2]}+T^{-}_{0}Q^{[+2]}.
\label{eqn:TQ_relation}
\end{equation}
The commutativity $[T_{j}(\mu),Q(\mu^{\prime})]=0$ for all values $\mu,\mu^{\prime} \in \CC$ and $j\in \ZZp$
enables us to operate on the level of operator spectra~
\footnote{Although the $Q$-operator associated with the transfer matrix is hermitian for $\lambda \in \RaR$, we keep the distinction 
between $Q$ and $\ol{Q}$ 
in Eq.~\eqref{eqn:TQ_relation} because in Sec.~\ref{sec:charge_densities} we will examine certain solutions of the $TQ$-equation for 
which the $Q$-functions are not real-valued.}
\footnote{For periodic boundary conditions, the 
$Q$-operator can exhibit singular behaviour, which can nevertheless be regularised, e.g.~by a boundary twist \cite{shortcut}. We 
stress however that in the following we do not construct the $Q$-operator explicitly~\cite{Korff05}, but instead only deal with 
well-behaved ratios of its eigenvalues.}.

The Bethe equations can be derived in an algebraic fashion from the $TQ$-equation \eqref{eqn:TQ_relation}, as the
condition for the cancellation of superficial poles in $T_{1}$ (see e.g.~\cite{shortcut}). Physically the 
Bethe equations specify the quantization conditions for momenta of excitations, and for a periodic chain of
length $\len$ they take the form
\begin{equation}\label{eqn:BE}
e^{\ii\,p(\lambda)\len}\prod_{k=1}^{M}S_{1,1}(\lambda-\lambda_{k})=-1,\quad \lambda=\lambda_1,\ldots,\lambda_M,
\end{equation}
where the two-particle scattering matrix $S_{1,1}(\lambda)$ is introduced in Appendix \ref{sec:AppendixA}.
The integer $M$ counts the number of flipped spins with respect to the totally polarized eigenstate, and $p(\lambda_j)$ is the 
momentum of a single magnon excitation which is related to the rapidity variable as
\begin{equation}
e^{\ii p(\lambda)}=\frac{\sin{(\lambda+\tfrac{\ii \eta}{2})}}{\sin{(\lambda-\tfrac{\ii \eta}{2})}}.
\end{equation}

In the non-deformed theory (i.e. for $q=1$), the Baxter $Q$-function (eigenvalue of the $Q$-operator on a Bethe state) is a 
degree $M$ polynomial whose roots are the Bethe roots $\lambda_{j}$. Under the quantum deformation, the $Q$-function becomes a product 
of trigonometric factors,
\begin{equation}
Q(\mu)=\prod_{j=1}^{M}\sin{(\mu-\lambda_{j})}.
\label{eqn:Q-operator}
\end{equation}

The Hirota equation \eqref{eqn:Hirota} permits a solution in terms of the $Q$-operator.
The linearity of Eq.~\eqref{eqn:TQ_relation} allows one to express
the higher-spin $T$-operators explicitly in terms of combinations of $Q$-operators~\cite{KLWZ97,Zabrodin97,Gromov09}
\begin{equation}
\frac{T^{+}_{j}}{T^{[j+1]}_{0}} =
Q^{[j+2]}\ol{Q}^{[-j]}\sum_{k=0}^{j}\frac{\zeta^{\len}_{j,k}}
{Q^{[2k-j]}Q^{[2k-j+2]}},
\label{eqn:solution_Hirota}
\end{equation}
where the scalar functions
\begin{equation}
\zeta_{j,k}(\mu) = \frac{\sin^{[2k-j+1]}(\mu)}{\sin^{[j+1]}(\mu)},
\label{eqn:zeta_functions}
\end{equation}
are defined as $\zeta^{\len}_{j,k}=T^{[2k-j+1]}_{0}/T^{[j+1]}_{0}$.

\subsection{String densities}
In the thermodynamic limit, solutions of the Bethe equations organize into regular patterns in the complex plane referred
to as (Bethe) strings. These are classified according to the
\emph{string hypothesis}~\cite{Takahashi71,Gaudin71,TakahashiBook,Volin12} and 
represent the thermodynamic particle content of the model. In the large-$\len $ limit many solutions of the Bethe equations become 
in fact indistinguishable and are characterized by the density distributions $\rho_{j}$ of the string centers. These densities
satisfy the set of integral equations known as string Bethe equations
\begin{equation}
\rho_{j}+\rhoh_{j}=a_{j}-a_{j,k}\star \rho_{k},
\label{eqn:string_Bethe_equations}
\end{equation}
adopting Einstein summation convention. Here $\rhoh_{j}$ are the densities of \emph{holes}, solutions of Eq.~\eqref{eqn:BE} which are 
\emph{not} Bethe roots, and the kernels $a_{j}$ and $a_{j,k}$, along with further details, can be found in 
Appendix~\ref{sec:AppendixA}.
As observed already in Bethe's seminal work~\cite{Bethe}, the strings are inherently local objects, and we will show in 
the next section that their density distributions are completely fixed by the local symmetries of the model.

\section{Thermodynamic spectra of local charges}
\label{sec:charge-string}
In this section we introduce the local conserved charges and derive their thermodynamic spectra. This leads to a direct
relationship to the string densities.
Again we focus our discussion on the $|\Delta|\geq 1$ regime, and refer the reader to Sec.~\ref{sec:gapless} for
subtleties related to the critical $|\Delta|<1$ regime.

\subsection{String-charge identification}
\label{sec:gapped}
We define a \emph{continuous} family of conserved operators $\XX_{j}(\mu)$~\cite{IMP15}
\begin{equation}
\XX_{j}(\mu) = \frac{1}{\len}\frac{1}{2\pi \ii}\pd_{\mu}\log
\frac{T^{+}_{j}(\mu)}{T^{[j+1]}_{0}(\mu)},\quad j=1,2,\ldots,
\label{eqn:X_log_form}
\end{equation}
where the scalar $T^{[j+1]}_{0}(\mu)$ provides a convenient normalization, cf. Eq.~\eqref{eqn:solution_Hirota}.
We will refer to these objects as the \emph{charges}, and the prefactor $N^{-1}$ ensures that their eigenvalues remain
finite in the thermodynamic limit. They become \emph{local} when $\mu$ is restricted to the physical region 
$\mathcal{P}_{\eta}$, and are hermitian for $\mu \in \RaR$.

To extract the spectra of $X_j$ we employ the Hirota equation and its solution in terms 
of the eigenvalues of the $Q$-operator. By inspecting the large-$\len$ behaviour of Eq.~\eqref{eqn:solution_Hirota}, we note that
for $k=j$ we have $\zeta_{j,j}(\mu)=1$, whereas $|\zeta_{j,k}(\mu)|<1$ for $k<j$, implying that only a single term
survives the large-$\len$ limit
\begin{equation}
\XX_{j}(\mu)=\frac{1}{\len}\frac{1}{2\pi \ii}\pd_{\mu}\log
\frac{Q^{[-j]}(\mu)}{Q^{[+j]}(\mu)},\qquad \mu \in \mathcal{P}_{\eta}.
\label{eqn:X_spectrum}
\end{equation}

Using the explicit form of the $Q$-function given by Eq.~\eqref{eqn:Q-operator} and the string hypothesis, expression 
Eq.~\eqref{eqn:X_spectrum} can be expanded over Bethe root densities
\begin{equation}
\XX_{j}(\mu)=\int_{-\pi/2}^{\pi/2}\dd \lambda\, \D_{j,k}(\mu-\lambda)\rho_{k}(\lambda)
\equiv \D_{j,k}\star \rho_{k},
\label{eqn:b_definition_gapped}
\end{equation}
where the kernels $\D_{j,k}$ can be neatly expressed in terms of the fused scattering matrices
\begin{equation}
\D_{j,k}(\lambda)=\sum_{i=1}^{k}\frac{1}{2\pi \ii}\pd_{\lambda}\log S_{j}(\lambda+(k+1-2i)\tfrac{\ii \eta}{2})=
\sum_{m=1}^{{\rm min}(j,k)}a_{|j-k|-1+2m}(\lambda).
\label{eqn:b_from_scattering_gapped}
\end{equation}
It is instructive to note that in Eq.~\eqref{eqn:b_from_scattering_gapped} the index $j$ designates the size of the
auxiliary spin, while the index $k$ pertains to the string type. We note that $G$ is the fundamental solution (``$\square G=\delta$", see Appendix \ref{sec:App_inversion}) of the operator $\square$,
the discrete d'Alembert operator introduced earlier in Eq.~\eqref{eqn:Laplace_gapped},
and so can be interpreted as a \emph{discrete Green's function} of the problem. Moreover, Eq.~\eqref{eqn:b_definition_gapped} shows that the $X_{j}$
manifestly comply with the \emph{additivity principle}, nicely exposing the particle nature of thermodynamic excitations.

The relation given by Eq.~\eqref{eqn:b_definition_gapped} is readily inverted upon application of the operator $\square$, 
as described in Appendix \ref{sec:AppendixA},
yielding the remarkable identity
\begin{equation}
\rho_{j} = \square \XX_{j}.
\label{eqn:Laplace_rho_X_gapped}
\end{equation}
Furthermore, with help of the string Bethe equations \eqref{eqn:string_Bethe_equations}, 
the hole distributions $\rhoh_{j}$ are also related to the charges in a local way
\begin{equation}
\rhoh_{j} = a_{j} - \XX^{+}_{j} - \XX^{-}_{j},
\label{eqn:rho_hole_to_Omega_gapped}
\end{equation}
recovering the previously known result from ref.~\cite{Ilievski_GGE}.

Let us stress that though $X_j$ are defined up to the gauge transformation of Eq.~\eqref{eqn:Hirota_gauge},
the string densities are independent of this.
Indeed, in the large-$\len$ limit the string densities can be lifted to the operator level and expressed in a
local gauge-invariant way in terms of the $Y$-operators from Eq.~\eqref{eqn:Y_to_T_relation} as
\begin{equation}
\rho_{j} = \frac{1}{\len}\frac{1}{4\pi \ii}\pd_{\mu}\log \frac{1+1/Y^{+}_{j}}{1+1/Y^{-}_{j}},\quad
\rhoh_{j} = \frac{1}{\len}\frac{1}{4\pi \ii}\pd_{\mu}\log \frac{1+Y^{-}_{j}}{1+Y^{+}_{j}}.
\end{equation}

Equation \eqref{eqn:Laplace_rho_X_gapped} will play a central role in the later discussion of quantum quenches.
It connects the charges, which can be explicitly evaluated on an initial state, to the density distributions of the strings
in the steady state.

\paragraph{Remark 1.}
Expanding the charge $\XX_{1}(\mu)$ about the origin recovers the well-known \textit{ultra-local} 
charges~\cite{Faddeev_arxiv,KorepinBook},
\begin{equation}
\XX_{1}(\mu)=\sum_{k=0}^{\infty}\frac{\mu^{k}}{k!}\XX^{(k)}_{1}.
\label{eqn:ultralocal_charges}
\end{equation}
In particular, the Heisenberg Hamiltonian Eq.~\eqref{eqn:Hamiltonian} is given by $X_{1}(0)$, up to an overall 
rescaling and shift of the spectrum. We want to stress nevertheless that, in principle, any charge
taken from the families $X_{j}$ can be legitimately considered as a Hamiltonian that is consistent with the same
two-particle scattering rule.
For characterization of states which encode local observables at equilibrium, the entire two-parametric family 
$\XX_{j}$ has to be accounted for on equal footing. While $\XX_{1}$ permits one to define Hamiltonians with
ultra-local Hamiltonian densities, generally the Hamiltonians contained in $\XX_{j}$ possess interactions which are of long 
range, but with exponentially decaying amplitudes (see ref.~\cite{IMP15}).
This weaker form of locality impacts the physics of local observables at the same level as the ultra-local charges.
Moreover, from the above consideration it is evident that it is preferable to operate with
the continuous representation rather than dealing with a countable discrete basis of charges of say 
Eq.~\eqref{eqn:ultralocal_charges}. We wish to emphasize that such a discrete basis has no obvious physical significance.

\paragraph{Remark 2.}
Let us mention a direct connection between locality as advocated in refs.~\cite{IMP15,Ilievski_GGE} 
and the large-volume behaviour of the $T$-system. Considering the large-$N$ limit of
\begin{equation}
\frac{T^{-}_{j}(\mu)}{T^{[-j-1]}_{0}(\mu)}
\frac{T^{+}_{j}(\mu)}{T^{[j+1]}_{0}(\mu)}=
1+Y_{j}(\mu),\qquad j \in \ZZp,\quad \mu \in \mathcal{P}_{\eta},
\label{eqn:inversion_identity}
\end{equation}
by taking into account Eq.~\eqref{eqn:solution_Hirota}, we observe that $1+Y_{j}(\mu)$ converges toward the \emph{identity}
operator as $\len \to \infty$. This provides an \emph{inversion relation}~\cite{IMP15}
which allows one to evaluate the logarithmic derivative in the definition of the charges Eq.~\eqref{eqn:X_log_form} as
\begin{equation}
\XX_{j}(\mu) \simeq \frac{1}{\len}\frac{1}{2\pi \ii}
\frac{T^{-}_{j}(\mu)}{T^{[-j-1]}_{0}(\mu)}
\pd_{\mu}
\frac{T^{+}_{j}(\mu)}{T^{[j+1]}_{0}(\mu)},
\label{eqn:X_lin_form}
\end{equation}
up to a correction which is subleading in system size $\len$. This result renders locality of $X_j$ manifest.
Moreover, locality of ref.~\cite{IMP15} can be understood as a corollary of the fusion rules among transfer matrices.

\paragraph{Remark 3.}
Recent studies of non-ergodic aspects of the Heisenberg spin chain uncovered a macroscopic family of \squote{hidden} local
conservation laws, referred to as the \squote{quasi-local} charges~\cite{ProsenPRL106,PI13,ProsenNPB14,Pereira14,IMP15}.
Curiously, quasi-locality in a weaker version appeared already in studies of non-ergodic Floquet dynamics of a quantum many-body
problem outside of conventional integrability~\cite{ProsenJPA98,ProsenPRE99}. Definition of quasi-locality does not make any direct
use of Bethe Ansatz related concepts but instead resorts to requiring \emph{extensive} $\sim \len$ scaling of
the Hilbert--Schmidt operator norm (a recent and more general formulation is presented in ref.~\cite{Doyon15}).
In the present work we evaded dealing with quasi-locality.
The upshot of our analysis is that quasi-locality is a manifest property of the large-volume scaling
of thermodynamic spectra of fused transfer operators. In addition, by providing a link to the
thermodynamic particle content of the spectrum which allows for a genuine local interpretation, the omission of the prefix 
\squote{quasi} is readily justified.
It remains an interesting open issue how the quasi-local conservation laws employed in 
refs.~\cite{ProsenPRL106,PI13,ProsenNPB14,Pereira14} which lie outside of the \squote{particle sector} (i.e.~objects pertaining to 
\emph{compact} representations of the symmetry algebra) can be connected to the framework presented in this article.

\paragraph{Remark 4.}

Finally, let us not forget the remaining local conserved operator $S^{z}$, allowing one to distinguish states
from symmetry multiplets which are characterized by same set of string densities.

\clearpage
\subsection{Gapless regime}
\label{sec:gapless}

For the gapless regime, modifications of the above formulation arise because the derivation of the string content is much more 
involved  compared to its gapped counterpart. This is a consequence of the exceptional spectral degeneracy which occurs when 
deformation parameter $q$ becomes a root of unity~\cite{TS72,TakahashiBook,KunibaTBA98}.
The string hypothesis is briefly summarized in Appendix~\ref{sec:AppendixA}. Two novel features to keep in mind are that (i) string 
configurations acquire an additional quantum number, the so-called string \emph{parity}, and (ii) that the number of distinct string 
types becomes \emph{finite}. Here we describe how these properties manifest on the level of the charges.

We proceed by retaining the structural form of the conservation laws from the gapped regime, i.e. make use of logarithmic derivatives 
of traces for the higher-spin monodromy operators.
Referring to the fact that strings are local objects, the number of (linearly) independent charges has to be in agreement with the 
number of distinct string types. In parallel to the stability condition for the strings (see Eq.~\eqref{eqn:stability_strings}), our 
task is now to derive an analogous condition for the gapless regime.
With this in mind, we first propose a three-parametric family of conserved operators $\XX_{(j,u)}(\mu)$ of the form
\begin{equation}
\XX_{(j,u)}(\mu)=\frac{1}{\len}\frac{1}{2\pi \ii}\pd_{\mu}
\log \frac{T^{+}_{(j,u)}(\mu)}
{T^{[j+1]}_{(0,u)}(\mu)},\qquad j=1,2,\ldots,\quad u=\pm 1,
\label{eqn:charges_gapless}
\end{equation}
writing shortly $T^{[\pm k]}_{(j,u)}=T_{j}(\mu \pm k\tfrac{\ii \gamma}{2}+(1-u)\tfrac{\ii \pi}{4})$ and restricting parameter $\mu$ to 
the physical strip
\begin{equation}
\mathcal{P}_{\gamma}\equiv \{\mu \in \CC|{\rm Im}(\mu) \in (-\tfrac{\ii \gamma}{2},\tfrac{\ii \gamma}{2})\}.
\label{eqn:physical_region_gapless}
\end{equation}
It has to be emphasized that even for $u=1$ conservation laws from Eq.~\eqref{eqn:charges_gapless} \emph{do not}
(in the $N\to \infty$ limit) automatically inherit locality from their gapped analogues 
under substitution $\eta\to \ii \gamma$ and $\mu \to \ii \mu$.
By keeping anisotropy parameter $\gamma$ fixed, only a subset of operators $\XX_{(j,u)}$, which will be
later on referred to as the \squote{\emph{charge content}}\footnote{Bethe root configurations in the thermodynamic limit for a given 
value of $\gamma$ will be referred to as the \squote{\emph{string content}}.}, is compatible with an extensive large-volume scaling as 
we shall readily demonstrate. By restricting the discussion to the roots of unity~\cite{TS72,KunibaTBA98},
$\gamma/\pi=\ell_{1}/\ell_{2}$ (cf. Appendix \ref{sec:AppendixA} for details) the total number of charges is always finite.

Specializing to the generic root of unity case, the number of string types is given by $m_{l}$ (cf. Appendix \ref{sec:AppendixA} for 
definitions and details). Remarkably however, the number of (linearly) independent charges is only $m_{l}-1$.
As we shall learn from the subsequent discussion, the reason is actually quite subtle and has to do with the \squote{truncation 
effect} of the $Y$-system hierarchy. Its physical consequence is that the so-called boundary pair of strings do not carry independent 
dynamical information.

Once the local conserved charges $X_{(j,u)}\in \mathcal{C}$ have been identified, we can proceed along the lines of the gapped 
scenario and rewrite their thermodynamic spectra $X_{(j,u)}$ in terms of Bethe roots distributions $\rho_k$
\begin{equation}
\XX_{(j,u)}(\mu)=\int_{-\infty}^{\infty} \dd \lambda\; \D_{(j,u),k}(\mu-\lambda)\rho_{k}(\lambda)
\equiv \D_{(j,u),k}\star \rho_{k},\quad \mu\in \mathcal{P}_{\gamma},
\label{eqn:X_spectrum_gapless}
\end{equation}
where $k$-th string which carries a pair of labels $(n_{k},v_{k})$ is now drawn from the string content for a particular
value of $\gamma$. The matrix Green function is determined by kernels $\D_{(j,u),(n,v)}$ which are provided it terms of scattering 
phases\footnote{Note that kernels $\D_{(j,u),
(n,v)}$ are well-defined objects even when a pair of labels $(n,v)$ is not taken from the string content.}
\begin{eqnarray}
\D_{(j,u),(n,v)}(\mu) &=& \sum_{i=1}^{n}
\frac{1}{2\pi \ii}\pd_{\mu}\log S_{j}(\mu+(n+1-2i)\tfrac{\ii \gamma}{2}+(1-uv)\tfrac{\ii \pi}{4})\\
\label{eqn:b_from_scattering_gapless}
&=& \sum_{m=1}^{{\rm min}(j,n)}a_{(|n-j|-1+2m,uv)}(\mu).
\end{eqnarray}

To find out which charge labels give rise to \emph{local} conservation laws we derive a stability condition based on 
the explicit solution of Hirota equation in terms of $Q$-functions.
Replacing $\lambda \to \ii \lambda$ and $\eta \to \ii \gamma$ in the scalars $\zeta_{j,k}(\mu)$ from
Eq.~\eqref{eqn:zeta_functions}, and incorporating the parity number by an appropriate 
$\ii \pi/2$ shift, the analysis reduces to study the moduli of
\begin{equation}
\zeta_{(j,u),k}(\mu)=\frac{\sinh^{[2k-j+1]}{(\mu+(1-u)\tfrac{\ii \pi}{4})}}
{\sinh^{[j+1]}{(\mu+(1-u)\tfrac{\ii \pi}{4})}},
\end{equation}
and requiring the \emph{stability condition}
\begin{equation}
\left|\zeta_{(j,u),k}(\mu)\right|<1,
\label{eqn:stability_condition}
\end{equation}
for all $k<j$, whereas by construction we have $\zeta_{(j,u),j}(\lambda)=1$.
Fulfilling condition \eqref{eqn:stability_condition} implies that conserved quantities with spectra given by expression 
\eqref{eqn:charges_gapless} attain additivity in the large-$N$ limit.
By inspection we find that condition \eqref{eqn:stability_condition} is fulfilled (for fixed $\gamma$) precisely
for $m_{l}-1$ pairs of labels $(j,u)$ which in effect determine $\XX_{(j,u)}$, i.e. the charge content of the theory.

\paragraph{Inverting matrix kernel $\D$.}
The problem of inverting relation \eqref{eqn:X_spectrum_gapless} is much more involved in comparison to
its gapped counterpart. Here we find ourselves forced to restrict the consideration to the zero magnetization sector,
and defer comments on polarized states to the conclusions (cf. Sec.~\ref{sec:conclusions}). This is due to a mismatch
between the number of strings $m_{l}$ and the number of charges $m_{l}-1$ which arises due to a redundancy linked to
the boundary string pair (associated with indices $m_{l}-1$ and $m_{l}$ in the standard ordering of ref.~\cite{TS72}).
In particular, since these two special string types scatter identically with respect to all other string types, and
\squote{inversely} among themselves~\cite{mirrorTBAI,mirrorTBAII}, i.e.~$S_{m_{l}-1}S_{m_{l}}=1$ and hence
$\D_{m_{l}-1,k}=-\D_{m_{l},k}$, the boundary strings are ascribed only a single charge.
To proceed, we restrict attention to states for which
the particle and hole distributions of the boundary string pair are related as
\begin{equation}
\rho_{m_{l}}=\rhoh_{m_{l}-1},\quad \rhoh_{m_{l}}=\rho_{m_{l}-1}.
\label{eqn:relation_boundary_strings}
\end{equation}
In appendix \ref{sec:App_magnetization} we show that this identification implies that such states have
zero magnetization\footnote{In principle there may exist states with zero magnetization for which 
Eq.~\eqref{eqn:relation_boundary_strings} is not true.}.

The redundancy arising from Eq.~\eqref{eqn:relation_boundary_strings} can be incorporated
by reducing the basis of $m_{l}$ densities $\rho_{j}$ to $m_{l}-1$ densities by modifying the last distribution at node $m_{l}-1$ 
according to the prescription
\begin{equation}
\rho_{m_{l-1}} \mapsto \widetilde{\rho}_{m_{l-1}}=\rho_{m_{l}-1}-\rho_{m_{l}}.
\label{eqn:modified_density}
\end{equation}
This allows us to define the matrix kernel $\D$ as a $(m_{l}-1)$-dimensional matrix of non-degenerate rank.
By inverting Eq.~\eqref{eqn:X_spectrum_gapless} we again arrive at an (almost) d'Alembert difference relation of the form
\begin{equation}
\rho_{j} = \square\XX_{j}.
\label{eqn:b_inverse_gapless}
\end{equation}
We remark that operator $\square$ now acquires an explicit dependence on the value of parameter $\gamma$ (which is
for clarity suppressed), which apart from determining its dimensions also introduces certain \squote{non-local} 
modifications. Before we state the explicit structure of $\square$ some extra clarifications are first in order.
In the gapless regime the auxiliary spin quantum numbers which label the charges no longer directly correspond to 
the ordering index $j$ in $X_{j}$ entering through Eq.~\eqref{eqn:b_inverse_gapless}, namely it
should not be confused with the sizes of auxiliary spin labels. It thus becomes crucial to decide about the ordering of charges 
explicitly. Here we assume the ordering of charges by \emph{increasing} spin label $j$. This conveniently renders
$\square$ of an almost-tridiagonal form in the linear space of densities $\rho_{j}$.

\paragraph{Principal roots of unity.}
The situation with simple roots of unity, given by $\ell_{1}=1$ and $\ell_{2}\equiv \ell$, becomes equivalent
to relation \eqref{eqn:Laplace_gapped} in the gapped case,
\begin{equation}
\square f_{j}=(s^{-1}-I\delta)_{j,k}\star f_{k},
\label{eqn:bbar_primitive}
\end{equation}
the only difference being that operators $\square$ and $I$ are of finite dimension $\ell-1$.
Relation \eqref{eqn:bbar_primitive} is a direct consequence of discrete d'Alembert identity \eqref{eqn:bulk_Laplace} and is 
determined by a single $s$-kernel reading $\mathcal{F}[s^{-1}](\omega)=2\cosh{(\tfrac{\pi}{2}p^{-1}_{0}\omega)}$.

\paragraph{Generic roots of unity.}
For \emph{non-principal} roots of unity, $\gamma/\pi=(\nu_{1},\nu_{2},\ldots,\nu_{l})$ for $l\geq 2$ (see
Appendix ~\ref{sec:AppendixA} and Eq.~\eqref{eqn:continued_fraction} for details), the $\gamma$-modified discrete d'Alembertian 
$\square$ component-wise reads
\begin{eqnarray}
\rho_{j} &=& (-1)^{i}\left(s^{-1}_{i+1}\star \XX_{j} - \XX_{j-1} - \XX_{j+1} \right),\quad m_{i}< j \leq m_{i+1}-1,\quad j\neq l-1,\\
\rho_{m_{i}} &=& (-1)^{i}\left(s^{-1}_{i+1}\star \XX_{m_{i}} - \XX_{m_{i-1}-1} - \XX_{m_{i}+1}\right),\quad i=1,\ldots l-1,\\
\widetilde{\rho}_{m_{l}-1} &=& (-1)^{l-1}\left(s^{-1}_{l}\star \XX_{m_{l}-1} - \XX_{m_{l}-2} - \XX_{m_{l-1}-1}\right),
\label{eqn:final_solution_gapless}
\end{eqnarray}
for $i=0,1,\ldots l-1$ and adopting the boundary condition $\XX_{0}\equiv 0$.

We wish to draw the reader's attention to three features which qualitatively differ from those of the gapped regime:
\begin{itemize}
\item A modification of the \squote{backward coupling} occurring at irregular indices pertaining to $l$-many exceptional nodes at 
positions $m_{i}$, for $i=1,2,\ldots l-1$, and finally $m_{l}-1$. These nodes can be interpreted as \squote{band edges}.
\item Nodes in the range $m_{i}\leq j \leq m_{i+1}-1$ are assigned convolution kernels $s_{i+1}$ which can be viewed
as various intrinsic length-scales present in the spectrum of the model.
\item Alternating overall signs $(-1)^{i}$ for distinct bands $m_{i} < j \leq m_{i+1}-1$.
\end{itemize}

A straightforward approach to prove Eq.~\eqref{eqn:final_solution_gapless} is to resort to Fourier representation of kernels
$a_{j}$ which transforms the main relation $\rho=\square\XX$ into a set of \emph{algebraic} equations, and subsequently employ
identities among scattering phase shifts which can be found in Appendix \ref{sec:AppendixA}.

\paragraph{Remark.}
This procedure is very reminiscent to the procedure used in the pioneering work on the
string hypothesis~\cite{TS72}, however it seems to us that the set of known kernel identities presented in
refs.~\cite{TS72,TakahashiBook} do not suffice to complete the proof due to presence of functions
$a_{(n,v)}$ in a typical expression for kernels $\D_{j,k}$ which actually do not directly involve the scattering data
for \squote{physical particles}.
Nonetheless, one can easily express the action of $s^{-1}_{i}\star$ on an arbitrary function $a_{(n,v)}$ in Fourier
space, i.e.~evaluate $\mathcal{F}[s^{-1}_{i}]\cdot \mathcal{F}[a_{(n,v)}]$, and observe cancellation of terms by
due to trigonometric addition formula $2\cosh{(\theta)}\sinh{(\psi)}=\sinh{(\psi+\theta)}+\sinh{(\psi-\theta)}$.

Some concrete examples for the non-principal roots of unity can be found in Appedix~\ref{sec:examples}.

\section{Evaluation of charge densities}
\label{sec:charge_densities}

Having established the identification \eqref{eqn:Laplace_rho_X_gapped} we are in position to immediately obtain
the string densities $\rho_{j}$ by evaluating the whole set of charges $X_{j}$ on a state $\ket{\Psi}$.
In the scope of a quantum quench, $\ket{\Psi}$ would be the initial state. 
The charges can be conveniently contracted with respect to $\ket{\Psi}$ by resorting to the form of Eq.~\eqref{eqn:X_lin_form}
\begin{equation}
\XX^{\Psi}_{j}(\mu)=\lim_{\len \to \infty}\frac{1}{\len}\frac{1}{2\pi \ii}
\bra{\Psi}
\frac{T^{-}_{j}(\mu)}{T^{[-j-1]}_{0}(\mu)}
\pd_{\mu}
\frac{T^{+}_{j}(\mu)}{T^{[j+1]}_{0}(\mu)}
\ket{\Psi}.
\label{eqn:X_psi0}
\end{equation}
Previously in the literature these objects have been customarily referred to as the \squote{generating functions} for the 
expectation values of the charges~\cite{FE13,FCEC14,WoutersPRL,Ilievski_GGE} with respect to state $\ket{\Psi}$\footnote{Here
we deal with a continuum of charges and hence the interpretation as the generating function does not make much sense.}.
Thanks to the Lax structure, an efficient computation is readily available with respect a large class of states, e.g. periodic matrix-
product states~\citep{FCEC14}. Hence for purely practical reasons we concentrate on states of the form
\begin{equation}
\ket{\Psi}=\ket{\psi}^{\otimes (\len/\len_{p})},
\end{equation}
where $\len_{p}\in \mathbb{N}$ pertains to the periodicity of the state.

By taking advantage of this product structure we employ the computational scheme which has been developed previously
in refs.~\cite{FE13,FCEC14,Ilievski_GGE}.
The main trick is to replace the logarithmic derivative of Eq.~\eqref{eqn:X_log_form} with the 
product form given by Eq.~\eqref{eqn:X_psi0}, where crucially the $\mu$-derivative is taken at the end of calculation,
i.e.~after contracting physical indices with respect to $\ket{\Psi}$.
Performing this however requires a small displacement of the spectral parameter, here denoted by $x$.
A local part of expression \eqref{eqn:X_psi0} is given by composite two-channel Lax operators
acting over $\CC^{2}\otimes \mathcal{V}_{j}\otimes \mathcal{V}_{j}$
\begin{equation}
\mathbb{L}_{j}(\mu,x)=\frac{L_{j}^{-}(\mu)}{L_{0}^{[-j-1]}(\mu)}\frac{L_{j}^{+}(\mu+x)}{L_{0}^{[j+1]}(\mu+x)}.
\label{eqn:two_channel_Lax}
\end{equation}
Computation of quantities $\XX^{\Psi}_{j}$ can be most elegantly achieved by means of the standard transfer matrix technique
via \emph{boundary partition functions}
\begin{equation}
Z^{\Psi}_{j}(\mu,x)=\lim_{\len \to \infty}\frac{1}{\len}{\rm Tr}_{\mathcal{V}_{j}\otimes \mathcal{V}_{j}}
\mathbb{T}^{\Psi}_{j}(\mu,x)^{\len/\len_{p}},
\label{eqn:boundary_partition_function}
\end{equation}
with
\begin{equation}
\mathbb{T}^{\Psi}_{j}(\mu,x) = \bra{\psi}\mathbb{L}^{(1)}_{j}(\mu,x)\cdots \mathbb{L}^{(\len_p)}_{j}(\mu,x)\ket{\psi},
\end{equation}
from where the charge distributions can be ultimately determined on taking the $x$-derivative at $x=0$,
\begin{equation}
\XX^{\Psi}_{j}(\mu)=-\ii\pd_{x}Z^{\Psi}_{j}(\mu,x)|_{x=0}.
\label{eqn:generating_function_definition}
\end{equation}
The large-$N$ behaviour of $Z^{\Psi}_{j}(\mu,x)$ is contained in the largest-modulus eigenvalue
$\Lambda^{\Psi}_{j}(\mu,x)$, which in the unperturbed limit obeys $\Lambda^{\Psi}_{j}(\mu,0)=1$,
by virtue of inversion relation \eqref{eqn:inversion_identity}.

\paragraph{Charge densities in the gapless regime.}
The procedure outlined above is equally valid in the gapless regime, the only difference being that the charge content 
becomes finite. All results can be computed along the lines of the gapped regime by making use of substitutions
$\lambda \to \ii \lambda$ and $\eta \to \ii \gamma$, and taking into account for the presence of negative parity.

\subsection{Explicit evaluation}
For a typical initial matrix-product state there exists an efficient numerical procedure to compute the charges,
introduced in~\cite{FE13} and extended in~\cite{FCEC14}. This is achieved by either directly referring to
Eq.~\eqref{eqn:generating_function_definition} or by using Jacobi's formula to rewrite
Eq.\eqref{eqn:generating_function_definition} as
\begin{equation}
\XX^{\Psi}_{j}(\mu)=\frac{-\ii}{\len_{p}}
\frac{{\rm Tr}\left({\rm Adj}(\mathbb{T}^{\Psi}_{j}(\mu,0)-1)\mathbb{D}^{\Psi}_{j}(\mu)\right)}
{{\rm Tr}\left({\rm Adj}(\mathbb{T}^{\Psi}_{j}(\mu,0)-1)\right)},
\label{eqn:Xeval}
\end{equation}
where $\mathbb{D}^{\Psi}_{j}(\mu)=\pd_{x}\mathbb{T}^{\Psi}_{j}(\mu,x)|_{x=0}$ and
the matrix coadjoint is given through $A\cdot{\rm Adj}(A)= \det{(A)}$.

\subsection{Analytic approach}
\label{sec:analytic}

In practice, explicitly evaluating Eq.~\eqref{eqn:Xeval} becomes a computational challenge already for relatively small values of $j$. In this section we show specific cases where this difficulty can be overcome. This is achieved by observing that for certain simple equilibrium states the set of functions $\eta_{j}=\rhoh_{j}/\rho_{j}$, which can be obtained through the string-charge duality relations Eqs.~\eqref{eqn:Laplace_rho_X_gapped},~\eqref{eqn:rho_hole_to_Omega_gapped}, satisfy the Y-system functional hierarchy Eq.~\eqref{eqn:Y-system}, which allows for a simple recursive solution to the problem. Moreover, by switching to the corresponding T-system we obtain a closed-form solution for the $\eta_{j}$ through the TQ-equation.
The density distributions $\rho_{j}$ then follow from the linear string Bethe equations \eqref{eqn:string_Bethe_equations}.
Let us stress however that it is an open question whether this procedure can be generalised to treat arbitrary initial states.

Specifically we consider two equilibrium states which have particularly simple \squote{representative}\footnote{Let us stress that these are not eigenstates, but rather
(from a quantum quench perspective) are in the basin of attraction of the equilibrium state.} states,
\begin{itemize}
\item \emph{N\'eel state} $\ket{\rm N}=\ket{\ua \da}^{\otimes \len/2}$, an eigenstate of the Ising limit $\Delta \to \infty$,
\item \emph{dimer state} $\ket{\rm D}=\tfrac{1}{\sqrt{2}}(\ket{\ua \da}-\ket{\da \ua})^{\otimes \len/2}$, a ground state of
Majumdar--Ghosh Hamiltonian~\cite{MG69},
\end{itemize}
on which the charges are straightforwardly evaluated. We remark that the N\'eel and dimer states have recently been found to permit
explicit evaluation of overlap coefficients with Bethe eigenstates, allowing for an exact implementation of the Quench Action 
method~\cite{CE13,Pozsgay_overlaps,WoutersPRL,Pozsgay_GGE,Brockmann14overlap,Mestyan15}. This time however we resort to 
a different technique.

We owe to stress that the solutions of Hirota equation \eqref{eqn:Hirota} we shall now discuss are \emph{distinct} from the 
canonical one of Sec. \ref{sec:Hirota}. We relax the condition that $T_0 = \phi^{-}=\ol{\phi}^{+}$, and consider a general
auxiliary linear problem for Eq.~\eqref{eqn:Hirota},
\begin{eqnarray}
\T_{j+1} \, \Q^{[j]} - \T^{-}_{j} \, \Q^{[j+2]} = \phi^{[j]} \, \ol{\Q}^{[-j-2]},\\
\T_{j-1} \, \ol{\Q}^{[-j-2]} - \T^{-}_{j} \, \ol{\Q}^{[-j]} = -\ol{\phi}^{[-j]} \, \Q^{[j]},
\label{eqn:Hirota_auxiliary_problem}
\end{eqnarray}
enabling to express its solution in the explicit form
\begin{equation}
\T_{k} = \T^{[-k]}_{0}\frac{\Q^{[k+1]}}{\Q^{[-k+1]}}+
\Q^{[k+1]}\ol{\Q}^{[-k-1]}\sum_{j=1}^{k}\frac{\phi^{[2j-k-1]}}
{\Q^{[2j-k-1]}\Q^{[2j-k+1]}}.
\label{eqn:T_solution_general}
\end{equation}
We wish to emphasize that the function $\Q$ here should not be confused with
Baxter's $Q$-operator given earlier by Eq.~\eqref{eqn:Q-operator} whose spectrum is a deformed polynomial which stores
the positions of the Bethe roots. On the contrary, here the analytic properties of $\Q$ encode the local physics of an 
equilibrium state.

As the $TQ$-equation is a second order difference equation, there generally exist two independent solutions to it.
For our purpose it is nonetheless sufficient to find only one solution.
The second solution can be in principle derived (modulo the addition of the first solution) by explicitly solving the quantum 
Wronskian condition~\cite{KLWZ97}.

We now present the explicit closed form solutions for the two states under consideration.

\subsubsection{Dimer state}

\paragraph{Isotropic point.}
At the isotropic point the auxiliary transfer operator
\begin{equation}
\mathbb{T}^{\rm D}_{j}(\lambda,x)=\bra{\rm D}\mathbb{L}^{(1)}_{j}(\lambda,x)\mathbb{L}^{(2)}_{j}(\lambda,x)\ket{\rm D},
\end{equation}
yields neat compact expressions for the first few charges,
\begin{equation}
\XX^{\rm D}_{1}(\lambda)=\frac{1}{2\pi}\frac{2\lambda^2+5}{4(\lambda^{2}+1)^{2}},\quad
\XX^{\rm D}_{2}(\lambda)=\frac{1}{2\pi}\frac{4(4\lambda^{2}+17)}{(4\lambda^{2}+9)^2},\quad
\XX^{\rm D}_{3}(\lambda)=\frac{1}{2\pi}\frac{3(2\lambda^{2}+13)}{4(\lambda^{2}+4)^{2}},
\end{equation}
while expressions pertaining to the charges of higher order shall be omitted here.
The corresponding expressions for $\eta$-functions are
\begin{equation}
\eta^{\rm D}_{1}(\lambda) = \frac{3\lambda^{2}}{1+\lambda^{2}},\quad
\eta^{\rm D}_{2}(\lambda) = \frac{32\lambda^{2}}{9+4\lambda^{2}},\quad {\rm etc.},
\end{equation}
and can be be encoded in $\T$-functions (modulo gauge freedom) simply as
\begin{equation}
\T^{\rm D}_{j}(\lambda)=(j+1)\lambda.
\end{equation}
The $\Q$-function for this particular case reads $\Q^{\rm D}(\lambda)=\lambda^{2}$. Moreover, here we were able to find the
other independent solution to Eq.~\eqref{eqn:TQ_relation}, $\Q^{\rm D}(\lambda)=-\ii/2$, which allows for a
determinant representation of $\T^{\rm D}_k$~\cite{Cherednik87,BR90,KLWZ97}.

\paragraph{Gapped case.}
Repeating the procedure in the gapped regime for an arbitrary value of $\eta$ one can extract the expressions for
the values of $\XX^{\rm D}_{1}(\lambda)$.
Their analytic form is quite cumbersome and so we omit them here.
The corresponding $\eta$-functions are somewhat simpler expressions and read e.g.
\begin{eqnarray}
1+\eta^{\rm D}_{1}(\lambda) &=& \frac{\cos{(4\lambda)}-\cosh{(2\eta)}}{\cos{(\lambda)}^{2}(\cos{(2\lambda)}-\cosh{(2\eta)})},\\
\eta^{\rm D}_{2}(\lambda) &=& -\frac{4\sin{(2\lambda)}^{2}(2\cos{(2\lambda)}+\cosh{(\eta)}+\cosh{(3\eta)})}
{(\cos{(2\lambda)}+\cosh{(\eta)})^{2}(\cos{(2\lambda)}-\cosh{(3\eta)})}.
\end{eqnarray}
Choosing $\T^{\rm D}_{1}(\lambda)=\sin{(2\lambda)}$ and setting the potential to
$\phi^{\rm D}(\lambda)=\sin{(\lambda+\tfrac{\ii \eta}{2})}\cos{(\lambda-\tfrac{\ii \eta}{2})}$, we obtain a set of
$\eta$-functions obeying the $Y$-system functional hierarchy compatible with the following set of $\T$-functions,
\begin{eqnarray}
\T^{\rm D}_{0}(\lambda) &=& \tfrac{1}{2}\sin{(2\lambda)},\\
\T^{\rm D}_{1}(\lambda) &=& \sin{(2\lambda)},\\
\T^{\rm D}_{2}(\lambda) &=& \tfrac{1}{2}\tan{(\lambda)}(3\cos{(2\lambda)}+\cosh{(2\eta)}+2),\\
\T^{\rm D}_{3}(\lambda) &=& \frac{\sin{(2\lambda)}(2\cos{(2\lambda)}+\cosh{(\eta)}+\cosh{(3\eta)})}
{\cos{(2\lambda)}+\cosh{(\eta)}},\quad {\rm etc.}.
\end{eqnarray}
The $\Q$-function is given by
\begin{equation}
\Q^{\rm D}(\lambda)=\cos{(\lambda-\tfrac{\ii\eta}{2})}.
\end{equation}

\subsubsection{N\'eel state}

\paragraph{Isotropic point.}
Here
\begin{equation}
\XX^{\rm N}_{1}(\lambda)=\frac{1}{2\pi}\frac{1}{2\lambda^2+1},\quad
\XX^{\rm N}_{2}(\lambda)=\frac{1}{2\pi}\frac{12}{12\lambda^{2}+19},\quad
\XX^{\rm N}_{3}(\lambda)=\frac{1}{2\pi}\frac{3\lambda^2+1}{2\lambda^4+7\lambda^2+2},
\end{equation}
and so forth, whence we readily calculate the first two $\eta$-functions
\begin{equation}
\eta^{\rm N}_{1}(\lambda) = \frac{\lambda^{2}(19+12\lambda^{2})}{(1+\lambda^{2})(1+4\lambda^{2})},\quad
\eta^{\rm N}_{2}(\lambda) = \frac{8(2\lambda^{2}+1)(2\lambda^{4}+7\lambda^{2}+2)}
{\lambda^{2}(\lambda^{2}+1)(4\lambda^{2}+9)}.
\end{equation}
The solution of the Hirota equation takes the form
\begin{equation}
\T^{\rm N}_{0}(\lambda) = \lambda,\quad
\T^{\rm N}_{1}(\lambda) = 2\lambda+\frac{1}{\lambda},\quad \Q^{\rm N}(\lambda) = 2\lambda.
\end{equation}

\paragraph{Gapped case.}
The first three initial charges take the form
\begin{small}
\begin{eqnarray}
\XX^{\rm N}_{1}(\lambda) &=& \frac{1}{2\pi}\frac{\sinh{(2\eta)}}{1-2\cos{(2\lambda)}+\cosh{(2\eta)}},\\
\XX^{\rm N}_{2}(\lambda) &=& \frac{1}{2\pi}\frac{2\sinh{(3\eta)}}{2\cosh{(3\eta)}+\cosh{(\eta)}-3\cos{(2\lambda)}},\\
\XX^{\rm N}_{3}(\lambda) &=& \!\!\frac{(2\pi)^{-1}\sinh{(4\eta)}(3\cos{(2\lambda)} - \! \cosh{(2\eta)}-2)}
{\cos{(2\lambda)}(3\cosh{(4\eta)}+ \! 2\cosh{(2\eta)}+3) - \! 2\cosh{(2\eta)}^{2}(\cosh{(2\eta)} + \! 2)- \! 2\cos{(4\lambda)}}.
\label{eqn:Neel_generating_explicit}
\end{eqnarray}
\end{small}
This gives
\begin{equation}
\eta^{\rm N}_{1}(\lambda)=
\frac{2\sin{(2\lambda)^{2}\left(2\cosh{(3\eta)}+\cosh{(\eta)}-3\cos{(2\lambda)}\right)}}
{(\cos{(2\lambda)}-\cosh{(\eta)})(\cos{(4\lambda)}-\cosh{4\eta})},
\end{equation}
while expressions for higher $\eta$-functions are suppressed here.
The solution of the Hirota equation is now cast in the form
\begin{eqnarray}
\T^{\rm N}_{0}(\lambda) &=& \tfrac{1}{2}\sin{(2\lambda)},\\
\T^{\rm N}_{1}(\lambda) &=& \tfrac{1}{2}\cot{(\lambda)}(1-2\cos{(2\lambda)}+\cosh{(2\eta)}),\\
\Q^{\rm N}(\lambda) &=& 2\sin{(\lambda)}.
\end{eqnarray}

\paragraph{Remark.}
The exact analytic from of the steady-state solution for the N\'eel quench problem has been presented
before in refs.~\cite{Brockmann14,Mestyan15} using the Quench Action approach~\cite{CE13}.
The authors of ref.~\cite{Brockmann14} already observed that the solution can be cast in the $Y$-system form, and 
in addition obtained an auxiliary function $\mathfrak{a}(\lambda)$ of the
Quantum Transfer Matrix method~\cite{Klumper93,Klumper02,GKS04}.
In this section we have shown that the latter is still a composite object, reducible in terms of auxiliary functions
$\Q$ and $\T_{0}$ as
\begin{equation}
\mathfrak{a}=\frac{\T_{0}^{-}\Q^{[+2]}}{\T^{+}_{0}\ol{\Q}^{[-2]}}.
\end{equation}
The observation that functions $\Q^{\rm N}$ and $\T^{\rm N}_{0}$ reproduce $\mathfrak{a}^{\rm N}$ for the N\'eel state found
in ref.~\cite{Brockmann14} (see also~\cite{Mestyan14}) represents a non-trivial compatibility check of the two approaches.

\section{Conclusions}
\label{sec:conclusions}

In this paper we have developed an explicit and transparent framework to describe the relationship between local symmetries
and equilibrium states, recently discovered in ref.~\cite{Ilievski_GGE}.
The charges in fact contain all information relevant to the steady-state expectation values of local observables,
obliviating the need to invoke statistical ensembles and a maximal entropy principle.
Any set of initial states, which are indistinguishable with respect to the charges, relax, via dephasing, to the same equilibrium 
state in the late-time limit. Presence of entropy merely reflects the fact that there are many microstates corresponding to the 
same macrostate, distinguishable only non-locally.

We have identified the charges (which are expressed explicitly over the local spin basis) with the string densities in an appealing 
form as a discrete wave equation \eqref{eqn:Laplace_rho_X_gapped}.
We exemplified this connection explicitly on a prototypical integrable quantum lattice model -- the anistropic Heisenberg spin-$1/2$ 
chain -- including both the gapped and the critical phases.
While in the gapped regime the discrete d'Alemebertian takes a purely local form, in the critical regime
it reduces to a finite-dimensional object and undergoes non-local modifications.

We presented the formalism in the language of fusion hierarchies. This provides a unified perspective, and
readily permits extensions to integrable lattice models based on higher-rank symmetry (super)algebras~\cite{KSZ08},
where each independent node on the $Y$-system lattice is assigned an individual charge.

A key application of the identification \eqref{eqn:Laplace_rho_X_gapped} is to address the situation of quantum 
quenches~\cite{CalabreseCardy,CalabresePRL106,Fioretto12,Calabrese12I,Calabrese12II,KormosPRB13,Sotiriadis14,Bertini14,Panfil15,
Collura15,Bonnes15}. We have presented a programme to access the steady state, which can be summarized with the following three-stage 
sequence
\begin{equation}
\ket{\Psi}\stackrel{{\rm (A)}}{\lra} \XX^{\Psi}_{j}(\lambda) \stackrel{{\rm (B)}}{\lra} \rho^{\Psi}_{j}(\lambda)
\stackrel{{\rm (C)}}{\lra} \expect{\mathcal{O}_{\rm loc}}.
\label{eqn:scheme}
\end{equation}
While steps $({\rm A})$ and ${(\rm B)}$ were a part of considerations in the present work, step $({\rm C})$ requires
invoking some extra tools and thus remains to be addressed in the future. Presently, the mapping $({\rm C})$ can
be readily implemented for the gapped and isotropic regimes (for applications see 
refs.~\cite{WoutersPRL,Pozsgay_GGE,Brockmann14,Mestyan15,Ilievski_GGE}) by employing formulae provided in
ref.~\cite{Mestyan14}, building on previous works~\cite{GKS04,Boos07,Trippe10}.

Further in this direction, the question of relaxation towards equilibrium is of central importance.
A powerful tool to investigate this is the Quench Action method, which allows one to track the finite-time evolution of local
correlators~\cite{CE13}. A recent study on the integrable Bose gas revealed a power-law decay in the
late-time dynamics~\cite{DeNardis15}. It would be interesting to address this issue in the framework of this article.

A distinguished property of the gapless regime is the appearance of an exceptional spectral degeneracy, which renders the
particle content finite. A subtle artefact of this reduction is the presence of a pair of boundary strings,
which cannot be distinguished on the level of charges.
Presently, we have only succeeded in resolving this boundary effect by restricting to a subset of states with zero
magnetization, while the general case of a polarized state remains a topic of future study.
Along the lines of the discussion in refs.~\cite{mirrorTBAI,mirrorTBAII}, for a polarized state
the corresponding $\eta$-functions of the boundary pair are expected to be related as $\eta_{m_{l}-1}\eta_{m_{l}} = e^{-\chi}$,
for some $\chi$. Such a constraint is however non-linear, and it remains an open question the matrix kernel $\D$ can
be amended in such a way to account for it.

We succeeded in casting two equilibrium states as closed-form solutions to quantum Hirota equation.
These states are however atypical, and general states lack such a compact description. It would be
interesting to realize a modified hierarchy of functional relations which would encompass all equilibrium states.

Finally, we wish to stress that $S^{z}$ and the family of charges $\XX_j$ do not in fact exhaust \emph{all} local
symmetries of the model at hand. In the gapless regime there exist additional local conserved operators associated with
non-compact highest-weight representations of
$\mathcal{U}_{q}(\mathfrak{sl}(2))$~\cite{ProsenPRL106,PI13,ProsenNPB14,Pereira14,Zadnik16}.
These charges are related to current-carrying non-equilibrium ensembles and are responsible for anomalous transport
properties, i.e.~the diverging DC conductivity~\cite{Zotos97,Zotos99,Benz05,ProsenPRL106,IP13}.
Including them in a general classification of steady states remains a prominent and challenging task and will be pursued in the 
ongoing research.

\acknowledgments

We are indebted to B. Wouters and J.-S. Caux for many valuable and inspiring discussions, and
collaborations on earlier projects at University of Amsterdam closely related to the content of this work.
In particular express our gratitude to M. Collura for collaboration during the initial stage of the project.
EI is grateful to M. Medenjak and T. Prosen for insightful conversations related to locality in
various contexts and acknowledges discussions with F. Essler. JDN acknowledges numerous discussions with M. Fagotti.
We are moreover thankful to T. Prosen and J.-S. Caux for their comments on the manuscript.
EI and EQ acknowledge support from the Foundation for Fundamental Research on Matter (FOM).

\clearpage
\appendix

\section{String hypothesis and Bethe equations for strings}
\label{sec:AppendixA}

\subsection{Two-particle scattering matrices}
\label{sec:scattering_matrices}

Here we systematically introduce the key concepts and tools needed for implementation of our programme.
We begin by considering a basic object of a quantum integrable theory, the two-particle scattering matrix.
Integrability of the model reflects the fact that any particle scattering events can be factorized
as a sequence of two-particle events.

The form of the scattering matrix is in fact related to the quantum $R$-matrix via Bethe Ansatz equations.
In the Heisenberg model, the $2$-particle scattering matrix admits the form
\begin{equation}
S(\lambda,\mu)=\frac{\sin{(\lambda - \mu - \tfrac{\ii \eta}{2})}}{\sin{(\lambda - \mu + \tfrac{\ii \eta}{2})}}.
\label{eqn:S-matrix}
\end{equation}
Notice that the class of so-called fundamental integrable models possess $S$-matrices which depend only on the \emph{difference}
of the particle's spectral parameters, i.e. $S(\lambda,\mu)=S(\lambda-\mu)$. The analogue of Eq.~\eqref{eqn:S-matrix}
at the isotropic point $\Delta = 1$ follows after taking the scaling limit $\lambda \to \lambda \eta$, and subsequently sending
$\eta \to 0$. This will bring the scattering matrix into the rational form.

\paragraph{Gapless regime.}
The elementary $2$-particle scattering matrix is readily obtained by taking the gapped counterpart Eq.~\eqref{eqn:S-matrix}
and applying the substitution $\lambda \to \ii \lambda$ and $\eta \to \ii \gamma$, in effect replacing factors $\sin{(\lambda)}$ with 
$\sinh{(\lambda)}$.

\subsection{String hypothesis}
\label{sec:string_hypothesis}

\paragraph{String hypothesis in the gapped regime.}
The string hypothesis asserts that the Bethe roots align into vertical patterns in the complex
rapidity plane~\cite{Takahashi71,Gaudin71,TS72,TakahashiBook}.
These formations are referred to as \textit{strings}. Physically, they describe \emph{bounds states} in the spectrum of the model
and represent thermodynamic particle excitations of a generic integrable lattice model.

Analysing the string content in the gapped regime of the Heisenberg chain shows that an infinite tower of string types emerge.
An $n$-string is composed of $n$ rapidities located on the real line at position $\lambda^{n}_{\alpha}$,
\begin{equation}
\{\lambda^{n,i}_{\alpha}\}\equiv \{\lambda^{n}_{\alpha}+(n+1-2i)\tfrac{\ii \eta}{2}|i=1,2,\ldots n\}.
\label{eqn:strings_gapped}
\end{equation}
Here index $\alpha$ enumerates different strings of length $n$, while the internal index $i$ runs over all Bethe roots inside
an individual string.

A scattering event with the string of length $j$ is given by a \emph{fused} scattering matrix
\begin{equation}
S_{j}(\lambda-\mu) = \frac{\sin{(\lambda - \mu - j\tfrac{\ii \eta}{2})}}{\sin{(\lambda - \mu + j\tfrac{\ii \eta}{2})}},
\end{equation}
where we identify $S(\lambda)\equiv S_{1}(\lambda)$. For convenience we additionally define the trivial scattering matrix $S_{0}=1$.

A product of subsequent scatterings in Bethe equations involving $M$ roots can be split as
\begin{equation}
\prod_{j=1}^{M}\longrightarrow \prod_{k=1}^{\infty}\prod_{\beta=1}^{M_{k}}\prod_{\alpha \in \{\lambda^{n,i}_{\alpha}\}},\quad
M=\sum_{k=1}^{\infty}kM_{k}.
\end{equation}
Then the \textit{string Bethe equations} among different sting types in a finite system of length $\len$ in
the logarithmic form read
\begin{equation}
\len \log S_{j}(\lambda^{j}_{\alpha})=2\pi I^{j}_{\alpha} + 
\log \prod_{k=1}^{\infty}\prod_{\beta=1}^{M_{k}}S_{j,k}(\lambda^{j}_{\alpha}-\lambda^{k}_{\beta}),
\label{eqn:Bethe_equations}
\end{equation}
where integer quantum numbers $I^{j}_{\alpha}$ are determined by fixing a branch of the logarithm.
A central property of scattering matrices $S_{j}(\lambda)$ is a functional identity
\begin{equation}
\frac{S_{j}(\lambda+\tfrac{\ii \eta}{2}) S_{j}(\lambda - \tfrac{\ii \eta}{2})}{S_{j-1}(\lambda)S_{j+1}(\lambda)} = 1,
\label{eqn:S_Laplace_eta}
\end{equation}
which in the logarithmic form becomes a \emph{discrete d'Alembert equation}.

In order to be able to describe any scattering event in the theory, we define a set of kernels $a_{j}(\lambda)$,
\begin{equation}
a_{j}(\lambda)= \frac{1}{2\pi \ii}\pd_{\lambda}\log S_{j}(\lambda),\quad j=1,2,\ldots
\label{eqn:aj_gapped}
\end{equation}
representing derivatives of the scattering phases which belong to individual strings.
Subsequently we prefer to leave the dependence on the spectral parameter $\lambda$ implicit.
Furthermore, we introduce the kernel $s(\lambda)$ as a solution to the equation~\cite{TakahashiBook}
\begin{equation}
a_{j}-s\star(a_{j-1} + a_{j+1})=0,\quad j>1,
\label{eqn:s_definition}
\end{equation}
whereas at $j=1$ we have $a_{1}=s+s\star a_{2}$.

Unit parameter shifts $\pm \tfrac{\ii \eta}{2}$ play an instrumental role.
The inverse of convolving with respect to $s$ can be understood as the deconvolution operator $s^{-1}\star$.
In particular, its action represents a symmetrized combination of spectral parameter shifts
\begin{equation}
(s^{-1}\star f)(\lambda)=\lim_{\epsilon\to0}\big( f(\lambda + \tfrac{\ii \eta}{2}-\ii \epsilon)+f(\lambda - \tfrac{\ii \eta}{2}+\ii \epsilon)\big),
\label{eqn:s_inverse_definition}
\end{equation}
for functions free of singularities inside the physical region $\mathcal{P}_{\eta}$.
For instance, applying $s^{-1}\star$ to Eq.~\eqref{eqn:s_definition} and using d'Alembertian yields $\square a_{j}=0$ for $j>1$, while $\square a_{1}=\delta$.

\paragraph{String hypothesis in the gapless regime.}
The above reasoning can be repeated in the critical regime $\Delta=\cos{(\gamma)}$.
The main (however quite a profound) distinction with respect to the gapped phase is that certain string types become 
prohibited~\cite{TS72}. Even worse, which strings are allowed now depends on the anisotropy parameter $\gamma$ is a quite dramatic 
way.

To analyse the so-called string content we follow the standard route presented in refs.~\cite{TS72,TakahashiBook} (see also 
\cite{KunibaTBA98}), while restricting our considerations to \emph{roots of unity},
\begin{equation}
\frac{\gamma}{\pi}=\frac{\ell_{1}}{\ell_{2}},
\label{eqn:root_of_unity}
\end{equation}
where $\ell_{1}<\ell_{2}$ are two co-prime integer numbers and $\ell_{2}>2$. Without making any further restrictions on $\ell_{1}$
and $\ell_{2}$, a set of points parametrized by Eq.~\eqref{eqn:root_of_unity} essentially densely covers the entire critical interval.

Following refs.~\cite{TS72,TakahashiBook} we adopt the terminated continued fraction representation of the anisotropy parameter,
\begin{equation}
\frac{\gamma}{\pi}\equiv p^{-1}_{0}=\frac{1}{\nu_{1}+\frac{1}{\nu_{2}+\frac{1}{\nu_{3}+\cdots}}},
\label{eqn:continued_fraction}
\end{equation}
which may be compactly written as $\gamma/\pi=(\nu_{1},\nu_{2},\ldots \nu_{l})$. Parameter $l\in \mathbb{N}$ can be
regarded as the degree of the root of unity, representing the numbers of distinct \squote{bands} in which the string particles
can be arrange to. We shall moreover borrow a sequence of $m$-numbers defined as
\begin{equation}
m_{0}=0,\quad m_{i}=\sum_{k=1}^{i}\nu_{k},\qquad i=1,\ldots l,
\end{equation}
which helps in determining the string content for given $\gamma$.
A short summary of the main ingredients of the string hypothesis in the gapless regime comprises:
\begin{itemize}
 \item Apart from the string length, an extra quantum number $v\in \{\pm 1\}$, called the string parity, arises.
Centers of negative-parity strings are displaced by $\ii\,p_0$ away from the real axis.
 \item The allowed strings are determined by the \textit{stability condition}
\begin{equation} 
 v \sin{(\gamma(n-j))}\sin{(\gamma j)}>0, \qquad j=1,2,\ldots n-1.
 \label{eqn:stability_strings}
\end{equation}
 \item The total number of distinct string types is $m_{l}$. Hence, the number of \squote{degrees of freedom} at roots of unity
 values of $\gamma$ in the thermodynamic limit is always \textit{finite}.
\end{itemize}

The allowed string lengths and parities are explicitly computable with aid of a sequence of auxiliary numbers 
${y_{i}}$~\cite{TS72,TakahashiBook},
\begin{equation}
y_{-1}=0,\quad y_{0}=1,\quad y_{1}=\nu_{1},\quad y_{i}=y_{i-2}+\nu_{i}y_{i-1},
\end{equation}
which allows to express
\begin{equation}
n_{j} = y_{i-1} + (j-m_{i})y_{i},\quad v_{j} = (-1)^{\lfloor (n_{j}-1)/p_{0} \rfloor}.
\label{eqn:gapless_length_parity}
\end{equation}
Another distinction of the gapless phase is that the set of $s$-kernels gets larger. Strings in the range
$m_{i-1}\leq j \leq m_{i}-1$ are associated with the kernel $s_{i}(\lambda)$.
For this purpose another sequence of auxiliary numbers $p_{i}$ is used,
\begin{equation}
p_{0}=\frac{\pi}{\gamma},\quad p_{1}=1,\quad p_{i}=p_{i-2}-p_{i-1}\nu_{i-1}.
\end{equation}
Scattering matrices are of the form
\begin{equation}
S_{j}(\lambda,\mu)\equiv S_{(n_{j},v_{j})}(\lambda,\mu)=
\frac{\sinh{(\lambda-\mu - n_{j}\tfrac{\ii \gamma}{2}+(1-v_{j})\tfrac{\ii \pi}{4})}}
{\sinh{(\lambda-\mu + n_{j}\tfrac{\ii \gamma}{2}+(1-v_{j})\tfrac{\ii \pi}{4})}},
\end{equation}
while the scattering between $j$-th and $k$-th type of strings is described by
\begin{equation}
S_{(n_{j},v_{j}),(n_{k},v_{k})} =
S_{(|n_{j}-n_{k}|,v_{j}v_{k})}S_{(n_{j}+n_{k},v_{j}v_{k})}
\prod_{m=1}^{{\rm min}(n_{j},n_{k})-1}S^{2}_{(|n_{j}-n_{k}|+2m,v_{j}v_{k})}.
\end{equation}
The corresponding kernels are given by
\begin{equation}
a_{j}(\lambda)=-\frac{1}{2\pi \ii}\pd_{\lambda}\log S_{j}(\lambda)=
\frac{1}{2\pi}\frac{2\sin{(\gamma q_j)}}{\cosh{(2\lambda)}+\cos{(\gamma q_j)}},\quad j=1,\ldots m_{l},
\end{equation}
where $q$-numbers $q_{i}$ (with $i=0,1,\ldots,l-1$) can be provided recursively
\begin{eqnarray}
q_{j} &=& (-1)^{i}(p_{i}-(j-m_{i})p_{i+1}),\quad m_{i}\leq j \leq m_{i+1}-1,\quad i=0,1,\ldots l-1, \\
q_{0} &=& p_{0},\quad q_{m_{l}}=(-1)^{l}p_{l}.
\label{eqn:q-numbers}
\end{eqnarray}
Fourier representations of $a_{j}$ are of particularly simple form,
\begin{equation}
\mathcal{F}[a_{j}](\omega)=\frac{\sinh{(q_{j}\tfrac{\gamma}{2}\omega)}}{\sinh{(\tfrac{\pi}{2}\omega)}},
\end{equation}
where in our convention Fourier transform and its inverse read
\begin{equation}
\mathcal{F}[f](\omega) = \int_{-\infty}^{\infty} \dd \lambda \; e^{-\ii \lambda \omega} f(\lambda),\quad
\mathcal{F}^{-1}[f](\lambda) = \int_{-\infty}^{\infty}\frac{\dd \lambda}{2\pi} e^{\ii \lambda \omega}f(\omega).
\label{eqn:Fourier_transform}
\end{equation}
Despite there is only $m_{l}$ physical strings, it is advantageous to define scattering phase shifts for the string types which 
\emph{do not} occur in the string content,
\begin{equation}
a_{(n,v)} = -\frac{1}{2\pi \ii}\pd_{\lambda}\log S_{(n,v)}(\lambda)=\frac{1}{2\pi}\frac{2\sin{(\gamma n)}}
{v \cosh{(2\lambda)}-\cos{(\gamma n)}}.
\end{equation}
Fourier space counterparts are obtained after exploiting $\pi$-periodicity in the imaginary direction,
performing an elementary contour integration around the path
\begin{equation}
[-\tau,\tau]\cup [\tau,\tau+\ii \tau]\cup \ii[\tau+\ii \tau,-\tau+\ii \tau]\cup [-\tau+\ii \tau,-\tau],
\end{equation}
while sending $\tau \to \infty$, using invariance under reflection $\lambda \mapsto -\lambda$, and finally
picking residua of $a_{(n,v)}\exp{(-\ii \omega \lambda)}$ on the imaginary interval $[0,\ii\pi]$. This yields
\begin{eqnarray}
\mathcal{F}[a_{(n,+)}](\omega) &=& \frac{\sinh{((\kappa_{+}-np^{-1}_{0})\tfrac{\pi}{2}\omega)}}{\sinh{(\tfrac{\pi}{2}\omega)}},
\quad \kappa_{+}=2 \left\lfloor \frac{n}{2p_{0}}\right\rfloor +1, \\
\mathcal{F}[a_{(n,-)}](\omega) &=& \frac{\sinh{((\kappa_{-}-np^{-1}_{0})\tfrac{\pi}{2}\omega)}}{\sinh{(\tfrac{\pi}{2}\omega)}},
\quad \kappa_{-}=2\left\lfloor \frac{n+p_{0}}{2p_{0}}\right\rfloor,
\end{eqnarray}
supplemented with $a_{(n,v)}=0$ when $n\gamma=\pi/2$. Above we introduced the \textit{mode numbers} $\kappa_{\pm}\in \ZZp$.
By virtue of the identity
\begin{equation}
S_{(n,v)}S_{(\ell_{2}-n,(-1)^{\ell_{1}}v)}=(-1)^{\ell_{1}},
\label{eqn:scattering_identity}
\end{equation}
one might exploit a \squote{shortening condition} for strings of lengths larger than $n> \lfloor \tfrac{\ell_{2}}{2}\rfloor$,
which allow to be interpreted as strings of length $n \to \ell_{2}-n$ if their parity gets transformed as
$v \to (-1)^{\ell_{1}}v$. Our calculations were made with the first option.

\subsection{String Bethe equations}
\label{sec:string_Bethe}

Under the assumption that $\lambda_{\alpha}-\lambda_{\beta}=\mathcal{O}(\len^{-1})$ for large $\len$, the set of quantum numbers 
$I^{j}_{\alpha}$ in Bethe equations \eqref{eqn:Bethe_equations} can be smoothly interpolated. This enables to introduce
particle (hole) densities by counting the number of Bethe roots (vacancies) which form a string of length $j$ on the
interval $[\lambda,\lambda+\dd \lambda]$. The spectrum can then be represented by an infinite set of smooth
root distributions $\rho_{j}(\lambda)$,
\begin{equation}
\rho_{j}+\rhoh_{j}=a_{j}-a_{j,k}\star \rho_{k}.
\label{eqn:string_Bethe_equations_A}
\end{equation}
This set of equations is referred to as the string Bethe equations for the densities.
Notice that each density $\rho_{j}$ is assigned a complementary variable $\rhoh_{j}$.
Here the scattering kernels $a_{j,k}$ are of the form
\begin{equation}
a_{j,k}(\lambda) = \frac{1}{2\pi \ii}\pd_{\lambda} \log S_{j,k}(\lambda).
\end{equation}

Bethe equations for the strings given by Eq.~\eqref{eqn:string_Bethe_equations_A} can be cast in a universal local form.
This can be achieved most elegantly by operating with a matrix kernel $(a+\kd)^{-1}$, which component-wise reads
\begin{equation}
(a+\kd)^{-1}_{j,k}=\kd_{j,k}\delta-sI_{j,k},
\end{equation}
with the incidence matrix $I$ reading
\begin{equation}
I_{j,k} = \kd_{j-1,k}+\kd_{j+1,k}.
\label{eqn:incidence_matrix_gapped}
\end{equation}
Here we use simultaneously the Dirac delta function $\delta(\lambda)$ and the Kronecker delta symbol $\kd_{i,j}$.
Convolving a function $f_{j}(\lambda)$ (analytic inside \emph{physical region} Eq.~\eqref{eqn:physical_region_gapped})
with $(a+\kd)^{-1}_{j,k}$ yields
\begin{equation}
(a+\kd)^{-1}_{j,k}\star f_{k}=f_{j}-s\star(f_{j-1}+f_{j+1}).
\end{equation}
Additionally, we have the identity
\begin{equation}
(a+\kd)^{-1}_{j,k}\star (a_{k,m}+\kd_{k,m}\delta)=\kd_{j,m}\delta.
\end{equation}
By operating with $(a+\kd)^{-1}\star$ on the raw form of string Bethe equations (cf. Eq.~\eqref{eqn:string_Bethe_equations_A}
produces a local (nearest-neighbour) coupled system of equations,
\begin{equation}
\rho_{j}+\rhoh_{j}=s\star (\rhoh_{j-1}+\rhoh_{j+1}).
\end{equation}

\paragraph{Gapless regime.}
In the gapless regime the Bethe equations for strings in the raw format get modified by the presence of
parity, resulting in a finite number of $m_{l}$ coupled equations~\cite{TS72,TakahashiBook}
\begin{equation}
{\rm sign}(q_j)(\rho_{j}+\rhoh_{j})=a_{j}-a_{j,k}\star \rho_{k}.
\end{equation}

\section{Derivation of the Green function}
\label{sec:App_inversion}

Here we provide some extra details on establishing the central identity \eqref{eqn:Laplace_rho_X_gapped}.
The convolution kernels $\D_{j,k}$, which determined the matrix Green function (cf. Eq.~\eqref{eqn:b_from_scattering_gapped}
in the main text), are directly expressible as linear combinations in terms of $a_{j}$.
Two notable symmetry properties of $\D$-kernels are
\begin{equation}
\D_{j,k}=\D_{k,j},\quad \D_{j,k}(\lambda)=\D_{j,k}(-\lambda).
\end{equation}
Similarly it can be shown that the kernels $a_{j,k}$ are composite objects and split as
\begin{equation}
a_{j,k}=\D_{j-1,k}+\D_{j+1,k}.
\label{eqn:a_to_b}
\end{equation}
Using the \squote{quasi-d'Alembertian} relation for kernels $a_{j,k}$,
\begin{equation}
a_{j,k}=s\star(a_{j-1,k}+\kd_{j-1,k}\delta)+s\star(a_{j+1,k}+\kd_{j+1,k}\delta),
\end{equation}
we arrive at a compact representation
\begin{equation}
\D_{j,k}=s \star (a_{j,k}+\kd_{j,k}\delta),
\label{eqn:b_kernel_definition}
\end{equation}
The fact that $\D$ is the Green's function of the wave operator $\square$ can be readily explicitly verified.
This amounts to show that
\begin{equation}
(\square \D)_{i,k}\equiv (s^{-1} - I\delta)_{i,j}\star \D_{j,k} = \kd_{i,k}\delta.
\label{eqn:b_kernel_gapped}
\end{equation}
with the boundary conditions $\D_{0,k}\equiv 0$ implicitly assumed. A brief calculation (using
Eq.~\eqref{eqn:b_kernel_definition}) shows
\begin{eqnarray}
[(s^{-1}-I\delta)\star \D]_{j,k} &=& \D^{+}_{j,k} + \D^{-}_{j,k} - (\D_{j-1,k} + \D_{j+1,k})\\
&=& (a_{j,k}+\kd_{j,k}\delta)-(\D_{j-1,k}+\D_{j+1,k})=\kd_{j,k}\delta.
\end{eqnarray}

\section{Simple examples}
\label{sec:examples}

\paragraph{Principal roots of unity.}
As an explicit example we treat a sequence of the so-called \emph{principal} roots of unity. They belong to values $\ell_{1}=1$ and
$\ell_{2}\geq 3$. Subsequently we put $p_{0}=\ell_{2}\to \ell$. Restriction to the principal points is rather standard in the
literature, the reason being that the set of non-linear integral equations which are formulated in the scople of Thermodynamic Bethe 
Ansatz techniques close at a finite level and in addition takes the simplest analytic 
form~\cite{TS72,KunibaTBA98,mirrorTBAI,mirrorTBAII}.

Auxiliary numbers are given by
\begin{equation}
q_{j}=\ell-j,\quad j=1,2,\ldots \ell-1,\qquad q_{\ell}=p_{1}=1,
\end{equation}
while the Bethe equations for the strings read
\begin{eqnarray}
\rho_{j} + \rhoh_{j} &=& s_{1} \star (\rhoh_{j-1} + \rhoh_{j+1}),\quad j=1,2,\ldots \ell-2,\\
\rho_{\ell-2} + \rhoh_{\ell-2} &=& s_{1}\star (\rhoh_{\ell-3} + \rhoh_{\ell-1} + \rho_{\ell}),\\
\rho_{\ell-1} + \rhoh_{\ell-1} &=& \rho_{\ell} + \rhoh_{\ell} = s_{1}\star \rhoh_{\ell-2}.
\end{eqnarray}
There are $\ell$-many distinct string types in the string content
\begin{equation}
(j,+),\quad j=1,2,\ldots \ell-1,\qquad \mbox{and }\quad (1,-).
\end{equation}
Strings with positive parity are ordered according to their increasing lengths $n_{j}=j$, whereas the only
negative-parity string is placed at the end. It is crucial to bare in mind that the last two strings (i.e. the boundary pair)
play a very special role. In particular, these two string scatter inversely
(cf. Eq.~\eqref{eqn:scattering_identity})
\begin{equation}
S_{(1,-)}S_{(\ell-1,+)}=-1,
\end{equation}
while with respect to the remaining strings one finds
\begin{eqnarray}
S_{(1,+),(1,+)} &=& S_{(1,-),(1,-)},\\
S_{(1,-),(j,+)} &=& S_{(\ell -1,+),(j,+)},\quad j=1,2,\ldots \ell-1.
\end{eqnarray}
From this analysis we can conclude that only one string from the boundary pair carries dynamical information, while the other 
is in this respect redundant and can be eliminated.
On the other hand, d'Alambertian relation among the \squote{bulk nodes} remains unaffected,
\begin{equation}
\frac{S_{j}(\lambda+\tfrac{\ii \gamma}{2})S_{j}(\lambda-\tfrac{\ii \gamma}{2})}{S_{j-1}(\lambda)S_{j+1}(\lambda)}=1,\quad
j=1,2,\ldots \ell-1,
\label{eqn:bulk_Laplace}
\end{equation}
Truncation effect yields
\begin{equation}
\frac{S_{\ell-1}(\lambda+\tfrac{\ii \gamma}{2})S_{\ell-2}(\lambda-\tfrac{\ii \gamma}{2})}
{S_{\ell-2}(\lambda)}=S_{\ell}=-1.
\end{equation}
D'Alembert identities among scattering matrices for strings at principal roots of unity
$\Delta=\cos{(\pi/\ell)}$ essentially indicate that this discrete set of points very closely resembles the situation
at a generic value of anisotropy in the gapped regime. The difference is merely visible as the \squote{truncation effect} which
gives rise to an exceptional structure at the boundary (see~\cite{TakahashiBook} and refs.~\cite{KunibaTBA98,mirrorTBAI,mirrorTBAII}).

\paragraph{Non-trivial example: roots $(\nu_{1},\nu_{2})$.}
The number of strings is here equal to $m_{2}=\nu_{1}+\nu_{2}$, with lengths given by
\begin{equation}
n_{j}=
\begin{cases}
j & 1 \leq j \leq \nu_{1}-1 \\
1+(j-\nu_{1})\nu_{1} & \nu_{1} \leq j \leq m_{2}-1 \\
\nu_{1} & j=m_{2}
\end{cases}.
\end{equation}
Parities should be assigned as $v_{1}=1$, $v_{\nu_{1}}=-1$, while for the remaining indices one should use prescription
given by Eq.~\eqref{eqn:gapless_length_parity}. There are only two relevant kernels, namely $s_{1}$ and $s_{2}$, which are determined 
in terms of $p$-numbers in accordance with auxiliary numbers
\begin{equation}
p_{0}=\frac{1+\nu_{1}\nu_{2}}{\nu_{2}},\quad p_{1}=1,\quad p_{2}=\frac{1}{\nu_{2}}.
\end{equation}
Bethe equations for densities read
\begin{eqnarray}
\rho_{j} + \rhoh_{j} &=& s_{1}\star (\rhoh_{j-1}+\rhoh_{j+1}),\quad 1\leq j \leq \nu_{1}-2,\\
\rho_{\nu_{1}-1} + \rhoh_{\nu_{1}-1} &=& s_{1}\star \rhoh_{\nu_{1}-2} + \tilde{s}_{1}\star \rhoh_{\nu_{1}-1}
-s_{2}\star \rhoh_{\nu_{1}},\\
\rho_{j} + \rhoh_{j} &=& s_{1}\star (\rhoh_{j-1} + \rhoh_{j+1}),\quad \nu_{1} \leq j \leq \nu_{1}+\nu_{2}-2,\\
\rho_{\nu_{1}+\nu_{2}-2} + \rhoh_{\nu_{1}+\nu_{2}-2} &=& s_{2}\star (\rhoh_{\nu_{1}+\nu_{2}-3} + \rhoh_{\nu_{1}+\nu_{2}-1})
+s_{2}\star \rho_{\nu_{1}+\nu_{2}},\\
\rho_{\nu_{1}+\nu_{2}-1}+ \rhoh_{\nu_{1}+\nu_{2}-1} &=& \rho_{\nu_{1}+\nu_{2}}+\rhoh_{\nu_{1}+\nu_{2}}
= s_{2}\star \rhoh_{\nu_{1}+\nu_{2}-2}.
\end{eqnarray}
The relation between the charge indices and their respective spin and parity labels can be explicitly encoded as
\begin{eqnarray}
X_{j} &=& X_{(j,+)},\quad j=1,2,\ldots \nu_{1}-1,\\
X_{m_{1}+m_{2}-j} &=& X_{(\ol{j},\ol{u})} ,\quad j=\nu_{1},\nu_{1}+1,\ldots,\nu_{1}+\nu_{2}-1,
\end{eqnarray}
where $m_{1}=\nu_{1}$, $m_{2}=\nu_{1}+\nu_{2}$, and where $(\ol{j},\ol{u})$ are \squote{mirror indices} defined by identifications
\begin{equation}
\ol{j} = \ell_{2}-j,\quad \ol{u} = (-1)^{\ell_{1}}u.
\end{equation}

Below we provide two concrete examples. For instance, the string content for $(\nu_1,\nu_2)=(2,2)$ ($p_0 = \tfrac{5}{2}$) is
\begin{equation}
\{(1,+),(1,-),(3,+),(2,+)\},
\label{eqn:content_(2,2)}
\end{equation}
whereas the corresponding charge content is
\begin{equation}
\{X_{(1,+)},X_{(2,+)},X_{(4,-)}\}.
\end{equation}
The last two strings, i.e. $(3,+)$ and $(2,+)$, constitute the boundary pair and do not carry an independent dynamical content,
meaning that their distributions are simultaneously fixed upon specifying $X_{(2,+)}$.
Let us consider another example, e.g. $(\nu_1,\nu_2)=(2,3)$ ($p_0 = \tfrac{7}{3}$). Here the string content reads
\begin{equation}
\{(1,+),(1,-),(3,+),(5,-),(2,+)\},
\end{equation}
while the charges are
\begin{equation}
\{X_{(1,+)},X_{(2,+)},X_{(4,-)},X_{(6,+)}\}.
\end{equation}

\section{Magnetization density sum rule}
\label{sec:App_magnetization}

\subsection{Gapped regime}
Given a state $\ket{\Psi}$, summing over all integrated density distributions should amount precisely to the magnetization density
\begin{equation}
m = \lim_{\len \to \infty}\len^{-1}\bra{\Psi}S^{z}\ket{\Psi}.
\label{eqn:magnetization_definition}
\end{equation}
By supposing that $\ket{\Psi}$ is characterized by densities $\rho_{j}$, Eq.~\eqref{eqn:magnetization_definition} is cast as
\begin{equation}
m=\left[\int_{-\pi/2}^{\pi/2}\dd \lambda\sum_{j=1}^{\infty}j\rho_{j}(\lambda)\right]-\frac{1}{2}.
\label{eqn:magnetization_density}
\end{equation}
To simplify the notation we write integrations over the fundamental period shortly as
$1\star f\equiv \int_{-\pi/2}^{\pi/2}\dd \lambda f(\lambda)$. We furthermore introduce the integrated particle distributions
$1\star \rho_{j}$ and the integrated charge densities $1\star \XX_{j}$. First we state a few useful identities,
\begin{equation}
1\star s = \frac{1}{2},\quad 1\star a_{j}=1,\quad 1\star s\star f = (1\star s)(1\star f)=\tfrac{1}{2}(1\star f).
\end{equation}
Le $\square^{(t)}$ designate a \emph{truncated} version of $\square$ by retaining only densities $\rho_{j}$
up to $j\leq t$, and charges $\XX_{j}(\mu)$ up to $j\leq t+1$. This allows us to approximate the density of magnetization
up to degree $t$ as $m_{t}=-\tfrac{1}{2}+\sum_{j=1}^{t}n_{j}(1\star\rho_{j})$. The exact magnetization density is obtained
from the limit $m=\lim_{t\to \infty}m_{t}$. By applying $1\star$ on $\square^{(t)}$ we readily obtain
\begin{equation}
m_{t}=[(t+1)(1\star \XX_{t})-t(1\star \XX_{t+1})]-\tfrac{1}{2},
\label{eqn:alpha_b}
\end{equation}
whence under the assumption that the limit $\lim_{t\to \infty}(1\star \XX_{t})=1\star \XX_{\infty}$ exists, the self-consistent 
solution of Eq.~\eqref{eqn:alpha_b} corresponds to large-$j$ limit of the integrated charge eigenvalue
$m=1\star \XX_{\infty}-\tfrac{1}{2}$.

\subsection{Gapless regime}
\paragraph{Principal roots.}
Here we show that the identification of the string densities for the boundary pair as in Eq.~\eqref{eqn:relation_boundary_strings}
implies vanishing magnetization, $m=0$. Now we employ the integrated variables,
\begin{equation}
\frakn_{j} \equiv 1\star \rho_{j},\quad \frakh_{j} \equiv 1\star \rhoh_{j},\quad \frakI_{j} \equiv 1\star \XX_{j},\quad
\fraka_{j} \equiv 1\star a_{j},\quad \fraka_{j,k} \equiv 1\star a_{j,k}.
\label{eqn:integrated_variables}
\end{equation}
As a quick non-trivial consistency check we derive the sum rule for the principal points ($p_{0}=\ell$).
By assuming that a state $\rho$ associated to $\ket{\Psi}$ is unpolarized, i.e.~$m=0$, we
imposing the identification of the boundary particle types in the form $\rho_{(1,-)}\equiv \rho_{\ell}=\ol{\rho}_{\ell-1}$,
and readily show that
\begin{equation}
m=-\frac{1}{2}+\frakh_{\ell-1}+\sum_{j=1}^{\ell-1}j\,\mathfrak{n}_{j} = 
\ell(\mathfrak{I}_{\ell-1}+\ol{\mathfrak{n}}_{\ell-1})-\frac{1}{2}
=\frac{\ell}{4}\fraka_{\ell-2}-\frac{1}{2}=0.
\end{equation}
In the above calculation we accounted for that fact that at the principal points the integrated convolution kernels obey
\begin{equation}
\fraka_{j}=\frac{\ell-j}{\ell},\quad j\leq \ell.
\end{equation}
using $\frakh_{\ell-1}=\tfrac{1}{4}\fraka_{\ell-2}-\frakI_{\ell-1}$ and
$\rhoh_{j}=a_{j}-s^{-1}\star \XX_{j}$ which is valid for nodes in the range $j=1,2,\ldots \ell-2$.

\paragraph{Generic roots of unity.}
The situation with generic roots of unity is more involved due to the presence of exceptional nodes.
It proves useful to split the entire contribution into the regular part $m_{\rm reg} = \sum_{j=1}^{m_{l}-2}n_{j}\frakn_{j}$,
and finally adding the contribution of the boundary nodes $m_{\rm b} = n_{m_{l}-1}\frakn_{m_{l}-1}+n_{m_{l}}\frakn_{m_{l}}$,
\begin{equation}
m=m_{\rm reg}+m_{\rm b}=\sum_{j=1}^{m_{l}}n_{j}\,\frakn_{j}=\ell_{2}\left(\frakh_{m_{l}-1}-(-1)^{l}
(\frakI_{m_{l-1}-1}-\frakI_{m_{l}-1})\right),
\end{equation}
The remaining integrated hole distribution which needs to be determined is $\frakh_{m_{l}-1}$. This one can be
obtained by combining raw string Bethe equations at node $m_{l}-2$,
\begin{eqnarray}
\frakh_{m_{l}-2} + {\rm sign}(q_{m_{l}-2})(\fraka_{m_{l}-2,m_{l}-1}+\fraka_{m_{l}-2,m_{l}})\frakh_{m_{l}-1} &=& \nonumber \\
{\rm sign}(q_{m_{l}-2})\Big[\fraka_{m_{l}-2}
-\fraka_{m_{l}-2,m_{l}-1}(\frakn_{m_{l}-1}-\frakn_{m_{l}})
&-& \sum_{k=1}^{m_{l}-2}\fraka_{m_{l}-2,k}\frakn_{k}
\Big]-\frakn_{m_{l}-2},
\end{eqnarray}
with the local Bethe equation found at node $m_{l}-1$,
\begin{equation}
\frakh_{m_{l}-1}=\tfrac{1}{4}\frakh_{m_{l}-2}-\tfrac{1}{2}(\frakn_{m_{l}-1}-\frakn_{m_{l}}),
\end{equation}
along with $\frakn_{m_{l}-1}-\frakn_{m_{l}}=(-1)^{l-1}(\frakI_{m_{l}-1}-\frakI_{m_{l}-2}-\frakI_{m_{l-1}-1})$ and the
integrated kernels
\begin{equation}
\fraka_{(n,+)}=1-\frac{n}{p_{0}}+2\left\lfloor \frac{n}{2p_{0}}\right\rfloor,\quad
\fraka_{(n,-)}=-\frac{n}{p_{0}}+2\left\lfloor \frac{n}{2p_{0}}+\frac{1}{2}\right\rfloor.
\label{eqn:integrated_kernels}
\end{equation}
Notice that by virtue of $m=0$ we have $\frakn_{m_{l}}=\frakh_{m_{l}-1}$.

\newpage
\bibliography{StringToCharge}
\bibliographystyle{ieeetr}

\end{document}